\documentclass{article}

\begin{document}
\title{Identity, Geometry, Permutation And The Spin-Statistics Theorem}
\author{M. J. York\\975 S. Eliseo Dr. \#9, Greenbrae, CA 94904, USA\\{\it email}: mikeyork@home.com}
\date{Nov 14, 1999}
\maketitle
\begin{abstract}
We examine historic formulations of the spin-statistics theorem from a point of view that involves no quantum field theory and distinguishes between the observable consequences and the ``symmetrization postulate''. In particular, we make a critical analysis of concepts of particle identity, state distinguishability and permutation, and  particle ``labels''. We discuss how to construct unique state vectors and the nature of the full state descriptions required for this -- in particular the elimination of arbitrary $2\pi$ rotations on fermion spin quantization frames and argue that the failure to do this renders the conventional symmetrization postulate (and previous ``proofs'' of it) at best {\em incomplete}.

We discuss particle permutation in a general way for any pairs of particles, whether identical or not, and make an essential distinction between exchange and pure permutation. We prove a revised symmetrization postulate that allows us to construct state vectors that are naturally symmetric under pure permutation, {\em for any spin}. The significance of particle labels (which, in the exchange operation, are not permuted along with other variables) is then that they stand in for any asymmetry (order dependence) that is present in the full state descriptions necessary for unique state vectors but not explicit in the regular state variables. {\em The exchange operation is then the physical transformation that reverses any asymmetry implicit in the labels}.

We point out a previously unremarked geometrical asymmetry between all pairs of particles that is present whenever we choose a common frame of reference. We show how this asymmetry affects the construction of permutation symmetric state vectors,  compute the exchange phase for various state vectors using different spin quantization frames, and prove the Pauli Exclusion Principle and its generalization to arbitrary spin.

\end{abstract}

\section{Introduction}

There has recently been speculation that it is possible to prove the spin-statistics theorem without recourse to relativistic field theory. Duck and Sudarshan \cite{Duck&Sud1} have provided an extensive overview of such proofs, claiming that a simple proof that introduces no new physical principle is not possible and exposing some of the flaws in various attempts that have been made. In ref. \cite{Duck&Sud2} they give a fuller account of the history of the theorem and attempts to understand it. (Although they limit themselves to proofs based on field theory or geometry and do not cover S-matrix proofs such as that of Stapp \cite{Stapp}.)

A common feature of many of the proofs they criticize seems to be the assumption of a particular geometric construction of the exchange operator. For two of the simplest of these --- in which particle exchange is defined to be a rotation by $\pi$ about an axis passing through their center of mass and perpendicular to the line joining their positions --- see Broyles \cite{Broyles} and Bacry \cite{Bacry}. Since, loosely speaking, rotating both particles by $\pi$ is equivalent to rotating one by $2\pi$ relative to the other, the conventional sign change for fermions is obtained.

The unsatisfactory nature of the proofs of Broyles and Bacry lies in the assumption that exchange is given by the same $+\pi$ rotation for both particles. It could equally, and, in this author's opinion, more reasonably, be argued that exchange is given by a $+\pi$ rotation for one particle (e.g. the ``first'') and the {\it inverse} rotation by $-\pi$ for the other (the ``second''). At least, in this case, the exchange operation is {\em physically} its own inverse. In this situation, there is no $2\pi$ relative rotation and no sign change.

But there is no freedom to {\em choose} the physical nature of the exchange operator in either way (symmetrically or otherwise). Rather, the physical nature of the exchange operator is determined, as we shall show, by any asymmetry (or order dependence) introduced in the choices made in defining the wave function --- in particular, the spin quantization frames. The observable exclusion rules follow from the redundancy implied by permutation, {\em whatever choices are made}.

A similar criticism can be applied to the field theory proofs. Feynman \cite{Feynman} used an example from QED to argue that exchange in the field theory proofs is also equivalent to a $2\pi$ rotation on one particle. By analogy with the geometric case, it should be possible to define exchange in the field theory context also in such a way that it implies no such rotation by judicious choices in specifying spin quantization frames. In other words, the commutativity or anti-commutativity of creation operators depends on whether or not they are {\em defined} in an order dependent way to create state vectors that are either order dependent or not.

It should now be apparent that, if one takes appropriate care with defining spin quantization frames, it is possible to define wave functions in a symmetric way, for which the exchange operator is the identity operator, instead of a rotation on one particle with respect to the other. Therefore, the spin-statistics theorem can no longer be accurately characterized simply by the symmetrization or anti-symmetrization of wave functions according to whether their spin is integer or half-integer (the ``symmetrization postulate''). The correct statement of the spin-statistics theorem lies purely with the observable exclusion rules --- which do not depend on such choices.

\subsection{The Historical Legacy}\label{sec:histover}
However, none of the previous ``proofs'' known to this author seems to address this critical issue. By relying on the notion that the exchange phase implicit in the symmetrization postulate determines the observable behavior, without further qualification regarding the hidden choices made in defining spin quantization frames in an order dependent way, they are all, at best, {\em incomplete}.

Although this author's claim is to have unraveled some of the confusion and oversights that historically have surrounded the spin-statistics theorem, he will readily admit that he, too, has often been confused. However, experience has taught him that his own confusion arose from the historical legacy rather than any intrinsic complexity. Furthermore, there seem to be as many differing opinions on various aspects of the differing claims to a proof as there are physicists with whom he has discussed the matter, even when they apparently agree on the conclusion (the ``symmetrization postulate''). For this reason, the proof given in this paper, although essentially very simple in concept, has been greatly expanded in order to anticipate many of the possible objections that have their origins in the historical confusion.

In the next section we shall present a quick and simplified version of how the Pauli Exclusion Principle comes about. We do this, in order to show how simple, in principle, the proof is. In subsequent sections we offer further explanation, in pedantic detail, of the relationship between permutation and exchange, a revised symmetrization postulate for pure permutation, and the intrinsic geometrical asymmetry in two-particle states. In the final section we also show how more general exclusion rules can be obtained.

For now, we shall attempt to expose the major sources of historic error and confusion. Historically, two simple mistakes of oversight are typically made:

\begin{enumerate}
\item If an exchange phase is to be meaningful, then the wave function must be a single-valued function of the variables that describe the two states and the relationships between them up to an overall, but arbitrary, absolute scalar factor common to the wave functions for {\it both} orderings. If neither wave function is {\em separately} ambiguous in its phase, then there can be no ambiguity in the relative phase that relates them. The usual claim that determination of the exchange phase requires some additional but subtle principle that is not explicit in the wave function therefore violates the single-valued requirement. This would be ok, except that it also means that the exchange phase is then undetermined, contrary to the purpose of introducing the new principle! In other words, such a claim that a new principle is necessary to determine the exchange phase is {\em self-contradictory}.

The fact that this simple observation has been mostly ignored (with few exceptions\footnote{See Mirman\cite{Mirman}, Leinaas and Myrheim\cite{Lein&Myr} and Berry and Robbins\cite{Berry&Rob}, all of whom recognize the importance of single-valued wave functions. However, they do not seem to appreciate the full significance of this for constructing permutation symmetric wave functions.}) for more than seventy years can be attributed to the (incorrect) beliefs that (a) there {\em must} be a sign change under re-ordering to explain the observable facts and (b) this sign change is not explicable {\em without} such an extra physical principle. But the fact remains, that single-valuedness neither allows nor requires such a new principle unless it involves new variables upon which the wave function must depend -- in which case such dependency will determine the exchange phase if the wavefunction is single-valued. {\em Given a complete set of variables, any order dependence in a single-valued wave function is inherent in the way it is defined and leaves no scope for ambiguity in the relative phase for different orderings (or new principles, unspecified by any observable variables, intended to resolve this fictitious ambiguity).} However, this is not a problem because, to explain the Pauli Exclusion Principle, it is sufficient to show only that it is {\em possible} to define an antisymmetric single-valued wave function for fermions, not that all fermionic wavefunctions must {\em necessarilly} be antisymmetric.

\item Typically there is usually an implicit assumption that a single common frame of reference can be chosen without introducing any asymmetry between the particles, although the present author does not recollect ever seeing this assumption made explicit. In fact, surprising though it may seem, such an assumption is false and cannot be made -- but the failure to realize this was the reason it was not understood where the fermionic sign change came from. Rather, there is an inherent geometrical asymmetry in two-particle states --- in particular in specifying the relative orientations of the spin quantization frames --- that persists {\em even when both particles have the same spin quantization frame} (though it is significant only for fermions). (Perhaps the simplest way to see this asymmetry is in the CM frame where, if one particle's azimuthal angle ($\phi_a$) lies in the range $0\leq \phi_a < \pi$ then the other's ($\phi_b = \phi_a + \pi$) lies in the range $\pi \leq \phi_b < 2\pi$ if measured in the same frame\footnote{If the reader is puzzled as to why this implies an asymmetry that can effect the choice of spin quantization frame, then clarity will hopefully be restored on reading sections \ref{sec:quick}, \ref{sec:spatasym} and \ref{sec:momspace}. For the present, we will simply point out that spin quantization frames must be specified relative to the position or momentum, or frame of reference in which we measure the position or momentum, if they are not to be ambiguous}.) A consequence of this asymmetry is that, if it is not correctly accounted for in the full state descriptions, then it necessarily introduces an order dependence in the spin quantization frames (even when they are apparently the same frame). It is this possible order dependence that enables us to construct a wave function for fermions that is antisymmetric and thereby prove the exclusion rule.
\end{enumerate}

The usual symptoms of these oversights are an unnecessary special treatment of identical particles, as though they have a distinguishing identity apart from their state variables, a failure to define single-valued wave functions and state vectors and a failure to distinguish pure permutation from any accompanying physical ``exchange'' transformation that might be implied by any order dependence in describing the individual states. 

By insisting on a complete set of state variables (and this includes the specification of the spin quantization frame as well as the quantum numbers of a state), sufficient to define single-valued state vectors, it is possible (and natural) to construct state vectors that are always permutation symmetric, whether for fermions or bosons, whether each individual state is a separate particle or a composite system, and whether identical or not, by eliminating any artificial and unnecessary order dependence in the individual state descriptions. Pairs of identical particles are then represented by a subspace of the larger (and permutation symmetric) space representing all paired states (single particle or composite) in general and require no special treatment. 

Typically, however, the construction of permutation symmetric and single-valued state vectors is not done. Rather, ambiguously asymmetric state vectors are used because they are defined in such a way that they are not single-valued. They can only be made single-valued, by allowing that they are order dependent (since the particle order is the only variable available to represent the missing details of the state description required for single-valuedness). But this order dependence cannot be derived theoretically until the geometrical asymmetry is taken into account.

The conventional minus sign for fermion exchange is therefore consequent on an order dependent method used to construct the wave functions (or the creation operators) when using a common spin quantization frame. But this is, of course, just what we need to derive exclusion rules. Without finding a way to construct asymmetric wave functions, we would not be able to compute any exclusion rules. It is a consequence of permutation invariance for composite systems that such order dependent (potentially anti-symmetric) wave functions exist for both fermions and bosons -- though in the latter case, they only occur with composite quantum numbers. The conventional fermion asymmetry and the Pauli rule are just the best known example.

In general, the exchange operator, is either nothing more than mere permutation (the symmetric case) or is indeed a physical transformation (e.g. a relative rotation of one particle's spin quantization frame by $2\pi$) because of an  order dependence in describing the composite state --- whether explicit or implicit.

The general method for deriving exclusion rules, whether we are dealing with fermions or bosons, is: (1) define single-valued permutation symmetric state vectors, (2) construct alternative single-valued but order dependent state vectors and compute their exchange phase by relating them to the permutation symmetric state vectors and (3) take the limit of identical quantum numbers.

In all cases, the observable exclusion rules are consequent purely upon the basic postulates of quantum mechanics and rotation and permutation invariance and the inherent geometrical asymmetry of two-particle states. No additional physical principle is required. They can be most conveniently expressed in one common exclusion rule for both fermions and bosons even though the ``statistics'' differ, by utilizing states of definite {\it composite} spin.  (The rule is that, when the individual particles are completely indistinguishable in all variables other than the third component of spin, then their composite spin must be even.) In other words, there is no fundamental difference between fermions and bosons other than their spins, even though their differing spins lead to very different observable behaviors for states of multiple identical particles.

\section{The Quick Version}\label{sec:quick}

This section outlines the essential features of this derivation of the conventional exchange phase in order to get the basic concepts over. The rest of the paper provides the details.

\subsection{Recapitulation}
The conventional argument about wave-function symmetry is that there exists an exchange operator $X$ such that if $\Psi(\alpha,\beta)$ is the wave function for two identical particles with states described by the sets of variables $\alpha$ and $\beta$, then 
\begin{eqnarray}
\Psi(\alpha,\beta) = X \Psi(\beta,\alpha)
 = X^2 \Psi(\alpha,\beta)
\end{eqnarray}
and hence the eigenvalues of $X$ are $\pm 1$.

However, the simple product wave function 
\begin{equation}
\Psi(\alpha,\beta) = \psi(\alpha) \psi(\beta) \label{eqn:symmwavefunc}
\end{equation}
will always give an eigenvalue $+1$, whereas the experimentally observed exclusion rules lead us to believe (although, as we have argued, this is in fact dependent on certain implicit additional conventions) that for certain particles, the eigenvalue of $X$ is $-1$.

The normal explanation of this relies on conjecturing that the particles have some additional ``identity'' that can be contained in the labels ``1'' or ``2'' to {\it distinguish} them. Hence the product wave-function may be symmetric or anti-symmetric under interchange of these labels:
\begin{equation}
X \psi^1(\alpha) \psi^2(\beta) = \psi^1(\beta) \psi^2(\alpha)
= \pm \psi^1(\alpha) \psi^2(\beta)
\end{equation}
and hence the wave functions
\begin{equation}
\Psi(\alpha,\beta) =  \psi^1(\alpha) \psi^2(\beta) \pm \psi^1(\beta) \psi^2(\alpha)
= \pm \Psi(\beta,\alpha)
\end{equation}
will be symmetric or antisymmetric under exchange.

It is then stated, that without some extra physical principle, such as relativistic field theory, it is not possible to know which eigenstate of X applies. The {\em symmetrization postulate} states that the eigenvalue of X is $(-1)^{2s}$ where $s$ is the spin of the identical particles.

\subsection{A Contrarian View}
It is our view that the missing ingredient involves no new physical principle, but simply the recognition of the true significance of the labels ``1'' and ``2'' in a previously unremarked physical asymmetry.

So why are the labels introduced in the first place? The reason is to introduce an extra variable that differentiates one wave function from the other. This is necessary only because of an {\em insufficiency} in the state descriptions $\alpha$ and $\beta$ to unambiguously define single-valued wave functions. The labels enable us to regain single-valued wave functions. If, on the other hand, $\alpha'$ and $\beta'$ are adequately unambiguous, then we can use them instead of $\alpha$ and $\beta$ and the labels can be dropped because they must be implicit in the difference between $\alpha$ and $\alpha'$ and between $\beta$ and $\beta'$.

Let us see how to describe this mathematically. The simplest way is to take the wave function
\begin{equation}
\Psi^{12}(\alpha,\beta) = \psi^1(\alpha) \psi^2(\beta)
\end{equation}
and note that it can also be written
\begin{equation}
\Psi^{12}(\alpha,\beta) = \Psi(\alpha',\beta') = \chi(\alpha') \chi(\beta') = \psi^1(\alpha) \psi^2(\beta)
\end{equation}
where the sets of variables $\alpha'$ and $\beta'$ differ from $\alpha$ and $\beta$ by the inclusion of the information contained in the ``labels'' $1$ and $2$. Clearly, since $\chi(\alpha')$ and $\chi(\beta')$ are both scalars, then $\Psi(\alpha',\beta')$ is exchange symmetric when the {\it complete} set of variables $\alpha'$ and $\beta'$ are exchanged:
\begin{equation}
\Psi(\beta',\alpha') = \chi(\beta') \chi(\alpha') = \chi(\alpha') \chi(\beta') = \Psi(\alpha',\beta')
\end{equation}
If the labels distinguish the two state descriptions then it must be possible to include them in the state variables to obtain single-valued functions. If not, they can be dropped. Either way, we can always define a symmetric wave function. If the labels are necessary then they must be present somewhere in the state variables that are required for the single-valued wave function. Historically it was not understood what the physical significance of these labels might be, hence this simple observation that it is always possible to define an exchange symmetric wave function {\em appeared} to contradict experiment and was ignored.

We are claiming that the ability to define symmetric wave functions is fundamental to quantum mechanics. Asymmetric (order dependent) wave functions can also be defined but only when the labels ``1'' and ``2'' have a physical significance that distinguishes the individual states.\footnote{In some ways, our position is very similar to that adopted by Mirman\cite{Mirman}. However we differ from his claim that ``the particles' identity prevents us from assigning any {\it physical} meaning to the indices'' since there are circumstances where the labels have a physical significance in distinguishability of the state descriptions, and therefore in distinguishability of the particles, even when the particles are otherwise identical. We would say, rather, that only {\it state indistinguishability} prevents us from assigning any physical meaning to the indices.} Even when the labels are present, exchange asymmetry arises only when we do not exchange the labels with the other variables (or, equivalently, when we exchange the labels only). When this is the case, the nature of the exchange operator and its eigenvalues {\it is determined by the physical relationship governing the distinguishing labels}.

A large part of the paper which follows is devoted to explaining and expanding on this assertion of physical significance. For the purposes of this quick proof, the reader is invited to accept the simplicity of the above argument --- that it is always possible to define exchange symmetric wave functions (even for fermions) by including sufficient variables to give a single-valued wave function for each separate particle --- and join us in looking for the physical significance of possible distinguishing labels when dealing with fermions.

\subsection{State Variables For Single-Valued Single Particle Wave Functions}
Since the existence of a unique exchange phase depends on the construction of wave functions for which the relative phase is uniquely determined by the state variables, let us look at the state variables involved in constructing such a single-valued wave function. 

In co-ordinate space, the significant variables that concern us for a single particle are the position vector $\mathbf{r}$, the particle spin $s$ and its third component $m$. It is important to remember that the spin is usually quantized along the z-axis which is related to $\mathbf{r}$ by the orientation of $\hat{\mathbf{r}}$ in the frame of reference. In other words, the spin quantization frame is tied to the vector $\mathbf{r}$, by its orientation in that frame. 

However, the specification of these variables is well-known to be insufficient to define a single-valued wave function. If we rotate the frame of reference by $2\pi$, all these variables ($\mathbf{r},s,m$) are unchanged; yet because the spin-quantization frame of reference has been rotated, the wave function changes its phase by $(-1)^{2s}$. To obtain a single-valued wave function, we must specify the angular co-ordinates of $\hat{\mathbf{r}}$ over a wider range of polar angles than those limited to the physical space (or, equivalently the rotation which takes $\hat{\mathbf{r}}$ into the z-axis of the frame of reference). Thus if, instead of $\mathbf{r}$ we use $r$, $\theta$ and $\phi$, then we have a single-valued wave-function $\chi(r,\theta,\phi,s,m)$ over the space $-\infty < \theta < \infty$ and $-\infty < \phi < \infty$, with the property that, for instance,
\begin{equation}
\chi(r,\theta,\phi+2\pi,s,m) = (-1)^{2s} \chi(r,\theta,\phi,s,m)
\end{equation}
when the rotation is about the z-axis. Note that if we had limited $\phi$ to the physical space ($0 \leq \phi < 2\pi$), then $\phi+2\pi$ would be equivalent to $\phi$ and we would not be able to distinguish the rotated wave function from the unrotated wave function. Hence, extending the range of $\theta,\phi$ is what enables us to obtain a single-valued wave function that distinguishes two frames related by a $2\pi$ rotation.

If, on the other hand, we had continued to use $\mathbf{r}$ without specifying the angular variables, then uniqueness of the relative phase would require us to specify a different wave function when the frame of reference is rotated by $2\pi$:
\begin{eqnarray}
\psi(\mathbf{r},s,m) & = & \chi(r,\theta,\phi,s,m)\\
\psi'(\mathbf{r},s,m) & = & \chi(r,\theta,\phi+2\pi,s,m) = (-1)^{2s} \psi(\mathbf{r},s,m)
\end{eqnarray}

Without distinguishing $\psi$ from $\psi'$, there is, therefore, a fundamental ambiguity in such wave functions. The same will be true for any other wave functions that depend only on variables that are unchanged by $2\pi$ rotations.

\subsection{Geometrical Asymmetry For Particle Pairs}

As we have already mentioned, there is, in a common frame of reference,  an implicit geometrical asymmetry in the relative orientation of one particle with respect to the other --- although this asymmetry does not necessarily translate into an asymmetry in the wave function until one considers the implications of linking the spin quantization frames to the frame used to measure the position vectors.

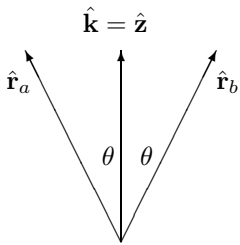
\begin{figure}[htbp]\begin{center}
\caption{\label{fig:bisect} Bisecting axis in co-ordinate space}
\setlength{\unitlength}{1in}
\begin{picture}(4,1.5)
\put(2.0,0.0){\vector(-1,2){0.5}}
\put(1.4,0.8){$\hat{\mathbf{r}}_a$}
\put(2.0,0.0){\vector(+1,2){0.5}}
\put(2.5,0.8){$\hat{\mathbf{r}}_b$}
\put(2.0,0.0){\vector(0,1){1.0}}
\put(1.8,1.1){$\hat{\mathbf{k}}=\hat{\mathbf{z}}$}

\put(1.9,0.4){$\theta$}
\put(2.1,0.4){$\theta$}
\end{picture}
\end{center}
\end{figure}

To see this asymmetry, note that the relative orientation can be specified by a rotation by $\pm \pi$ about the axis $\hat{\mathbf{k}}$ which bisects their position vectors (in co-ordinate space) (see fig. \ref{fig:bisect}) or which bisects their momentum vectors (in momentum space). For the rest of this section we shall work in co-ordinate space. If the rotation is chosen to be $+\pi$ for the orientation of $\mathbf{r}_b$ relative to $\mathbf{r}_a$ then it is $-\pi$ for the orientation of $\mathbf{r}_a$ relative to $\mathbf{r}_b$, since the latter must be the inverse of the former if the azimuthal angles in the plane perpendicular to the axis of rotation are to be swapped unchanged. Since each rotation is the inverse of the other, they cannot be the same and therefore are distinguishable. If one rotation is clockwise, the other must be counter-clockwise. 

We shall go into more detail about this asymmetry in section \ref{sec:spatasym}.

For the present discussion, all we need to know is that a single-valued two-particle wave function for arbitrary spin in a common frame of reference requires the unique specification of the relative orientation of one particle to the other. If this relative orientation was ambiguous, then the orientation of one particle's spin quantization frame relative to the other would be known only up to an arbitrary $2\pi$ rotation and therefore the wave function might be undetermined up to a sign change.

Clearly this unique specification can be done using the unambiguous variables ($r,\theta,\phi,s,m$) for each particle since the angular co-ordinates fix the relative orientation of $\mathbf{r}_a$ and $\mathbf{r}_b$. If, instead we were to use the ambiguous variables ($\mathbf{r},s,m$) we would need some other method, such as additional labels ``1'' and ``2'' to fix the relative orientation of ``1'' to ``2''. In this case, as we shall show, the relative orientation of $\mathbf{r}_a$ to $\mathbf{r}_b$ is reversed by the exchange --- which is equivalent to rotating one particle's frame of reference by $\pm 2\pi$. 

If this relative orientation is indeterminate, then the wave function will not be single-valued and the exchange phase will also be indeterminate.

Note that the two methods indicated for specifying the relative orientation are mutually exclusive. If the relative orientations of the particles are defined both before and after exchange by their labels ``1'' and ``2'', then they cannot also be defined by their angular co-ordinates, and {\it vice versa}.

To see this, consider a frame of reference in which the z-axis bisects the position vectors associated with the individual particles (fig. \ref{fig:bisect}). Using the angular co-ordinates to specify the relative orientations gives
\begin{equation}
\Delta_{ab} = \phi_b - \phi_a = \pm\pi
\end{equation}
whereas, using particle ordering to specify the relative orientations gives
\begin{equation}\label{eqn:del12def}
\Delta_{12} = \phi_2 - \phi_1 = \pm\pi
\end{equation}
and we see that it is possible to choose fixed values for either $\Delta_{12}$ or $\Delta_{ab}$ but not {\it both}. If we fix $\Delta_{12} = +\pi$, then we have $\Delta_{ab} = +\pi$ when $a=1$, but $\Delta_{ab} = -\pi$ when $b=1$.

The implicit distinction between ``1'' and ``2'' that we have specified in eqn. \ref{eqn:del12def} is equivalent to:
\begin{eqnarray}
\psi^{12}(\mathbf{r}_a,s_a,m_a,\mathbf{r}_b,s_b,m_b) = \chi(r_a,\theta,\phi^1_a,s_a,m_a) \chi(r_b,\theta,\phi^2_b,s_b,m_b)\nonumber\\
 = \chi(r_a,\theta,\phi^1_a,s_a,m_a) \chi(r_b,\theta,\phi^1_a+\pi,s_b,m_b)
\end{eqnarray}
where the condition $\phi^2_b = \phi^1_a + \pi$ is necessary for the unique specification of the relative orientations. Applying the exchange operator in such a way as to maintain $\phi^1_b = \phi^2_b = \phi^1_a + \pi$ unchanged, we find
\begin{eqnarray}\label{eqn:quickexchange}
X \psi^{12}(\mathbf{r}_a,s_a,m_a,\mathbf{r}_b,s_b,m_b) = \psi^{12}(\mathbf{r}_b,s_b,m_b,\mathbf{r}_a,s_a,m_a)\nonumber\\
 = \chi(r_b,\theta,\phi^1_b,s_b,m_b) \chi(r_a,\theta,\phi^2_a,s_a,m_a)\nonumber\\
 = \chi(r_b,\theta,\phi^1_b,s_b,m_b) \chi(r_a,\theta,\phi^1_b+\pi,s_a,m_a)\nonumber\\
 = \chi(r_b,\theta,\phi^1_a+\pi,s_b,m_b) \chi(r_a,\theta,\phi^1_a+2\pi,s_a,m_a)\nonumber\\
 = (-1)^{2s_a}\ \psi^{12}(\mathbf{r}_a,s_a,m_a,\mathbf{r}_b,s_b,m_b)
\end{eqnarray}
Alternatively, the interchange $1\leftrightarrow2$, keeping $\phi^2_a = \phi^1_a$ fixed is equivalent to replacing $\phi^2_b$ by $\phi^1_b = \phi^2_b-2\pi$ and results in an exchange eigenvalue of $(-1)^{2s_b}$. Either way, for identical particles ($s_a=s_b=s$), the eigenvalue of $X$ for these order-dependent wave-functions is $(-1)^{2s}$. And in the limit that $s_a=s_b=s$, $m_a=m_b=m$ and $\mathbf{r}_a=\mathbf{r}_b=\mathbf{r}$, we find the wave function $\psi^{12}(\mathbf{r},s,m,\mathbf{r},s,m)$ vanishes for half-integer $s$, which is exactly the Pauli principle.

However, there are two significant differences from conventional proofs. First we have shown that the anti-symmetric construction is not the only one. It just happens to be a convenient way to illustrate the exclusion rule implicit in the permutation symmetry. Second, we have shown how to compute the exchange phase in the anti-symmetric case, directly from the symmetric case. The symmetric (order independent) wave function, which uses fixed angular co-ordinates that are not affected by the exchange, coexists with the asymmetric (order dependent) wave function and their relationship is mathematically defined by their method of construction. 

The reader is now invited to read the rest of this paper for a more detailed explanation and to satisfy themselves that no smoke or mirrors are involved.

\section{Summary Of The Detailed Proof}
It has been the author's experience that much of what follows in this paper may appear at times to be either trivially obvious or apparently incorrect in different places for different readers, simply because of the confusing historical legacy. The purpose of this section is to expand on the introduction and thereby provide a gentle introduction to the method we use to cut through the confusion --- leaving the detail till later. Whenever the reader is tempted to react to a particular statement with the belief that it is obviously trivial or obviously wrong, given what they think they already know about the spin-statistics theorem and its assumed equivalence to the conventional symmetrization postulate, they are requested to ponder the possibility that it is not so obvious to others or that maybe what they already ``know'' might not be so simple as it appears.

The method we shall adopt and the key observations can be summarized in the following steps:

\begin{enumerate}
\item First of all, we look at particle identity and distinguishability. Identity rests in specific restrictions on some of the allowed quantum numbers of the state, such as the rest mass and the charge and has no connection to particle ``labels''. 

\item Allowed states of a specific particle are simply a special case of general quantum mechanical states representing either individual particles or composite systems (in which the constituent particles are indeterminate and only the composite quantum numbers matter). All discussion of particle exchange symmetry and its consequences, in this paper, applies equally well to composite systems in which we concern ourselves only with the composite quantum numbers.

\item However, the identity of the particles and the indistinguishability of their state descriptions are not the same. Although indistinguishable particles must be identical, the converse is not true. Just as two identical coins can be tossed in a way that distinguishes them, identical particle states can be described in ways that distinguish them. Indistinguishability of two states therefore requires indistinguishability in all variables that are required for a complete description of the state.

\item We then consider the significance of permutation symmetry in describing physical states. We argue that because permutation invariance is sufficient to determine the reduction in allowed states of identical tossed coins, then the same should be true for identical quantum-mechanical particles. Similarly, just as it is possible to describe states of paired coins in an order-free way, paired quantum-mechanical states are also capable of order-free description.

\item We then consider the significance of the well-known phase ambiguity in quantum-mechanical state vectors for states with the same quantum numbers. We argue the necessity of a {\em physical completeness} axiom by which it is possible to choose unique state vectors that are unchanged by non-physical transformations. This requires state descriptions that are physically complete in more than just their quantum numbers. We show what such physical completeness entails in the context of spin quantization frames.

\item We discuss the distinction between pure permutation and ``exchange''. Permutation involves a re-ordering of the sets of variables describing two states. When these individual state descriptions are complete and order independent, then their re-ordering is strictly a permutation. We prove a {\em revised symmetrization postulate} in which permutation of order independent and physically complete state descriptions cannot, by itself, change the phase of a single-valued state vector because it is not a physically significant transformation.

\item The exchange operation differs from pure permutation only when there is an additional distinguishability which is not permuted with the other state variables (that is, where exchange is equivalent to {\it partial} permutation). In such a case, exchange is also equivalent to the interchange of distinguishing features. Since this may involve a physically significant change in the individual state descriptions, it may result in a change of phase.

\item There exists a fundamental asymmetry in two-particle states between the relative orientations of each particle (whether the orientation vector is the position vector, the momentum vector, the spin quantization axis or some other vector) which arises because the rotation which takes one particle's orientation vector into that of the other is physically distinguishable from its inverse.

\item This has implications for the physical completeness and order independence of state descriptions for particle pairs and hence for choosing a permutation symmetric state vector. We show how to construct such a state vector using independent frames of reference and then also show how to relate them to a common frame. 

\item  We show how to construct different state vectors using different methods for choosing the spin quantization frames. Using the revised symmetrization postulate, we show how to compute the exchange phase in each case.

\item The fact that we can construct exchange symmetric or exchange asymmetric state vectors does not affect the observable behavior, since ---as long as they are single-valued --- either construction is uniquely related to the other. In the limit of identical individual quantum numbers, the asymmetric case will tell us the exclusion rules, although the underlying permutation symmetry is present in both cases.

\item Although the Pauli Principle is a convenient summary of the exclusion rules for fermions, the conventional exchange phase, by itself, says nothing about excluded states for bosons. More general rules for both fermions and bosons, can be expressed in terms of combined angular momentum\footnote{See, for example, Rose\cite{Rose}, chapter 12}. We show that, in a state in which all other quantum numbers are identical, the individual spins for the allowed states must combine such that the total spin quantum number $S$ is always even whether we have fermions or bosons (states of odd $S$ are forbidden). We also prove a more general rule that, in the Center of Mass frame, states of odd $L+S$ are forbidden.

\end{enumerate}

\section{Identity, Distinguishability And Permutation Invariance}	
\label{sec:PI}

The concept of identical particle exchange, is, by definition, one that applies to identical particles only. However, the more general principle of permutation of individual state descriptions is not limited to identical particles or even to single particles of any description, since the states being permuted may be composite. In this section, we will explore the distinction and the connections between exchange and permutation and between identity and distinguishability.

We shall also look at the idea of {\it permutation invariance} of physical states of any sort, whether particles or tossed coins. We argue that concepts of particle identity are essentially irrelevant to understanding permutation invariance and that it is the identity of all individual quantum numbers, not just particle type that results in the observable behavior expressed in exclusion rules.

\subsection{What Is Identity Anyway?}

The most common use of ``identity'' in quantum mechanics is the concept of particle ``type'' or ``character''. When we say that two particles are ``identical'' we mean they have the same {\em type}. By this we mean that there is a certain set of attributes (fixed quantum numbers such as mass, charge, etc) that are peculiar to that type of particle.

Classically, however, there is another common meaning of ``identity''. When we follow the trajectory of a {\em single} particle through space, we use the term ``identity'' to refer to a specific particle in which each successive observation of position is that of the {\em same} particle. In this sense the identity is unique to that specific particle and not just its type. Quantum mechanically, however, this notion of a single particle's unique identity is meaningless.

Any notion that particle ``labels'' distinguish this ``unique identity'' of each particle independently of their state variables contradicts the basic properties of observations in quantum mechanics.

To see this, suppose we make two successive observations. In each case we detect a system described by a single object with attributes $\sigma$ at positions $\mathbf{r}$ and $\mathbf{r} + d\mathbf{r}$. We might, by analogy with the classical situation, naturally tend to assume that we have detected the {\em same} object twice. But suppose the second observation gives a different set of attributes $\sigma' \ne \sigma$. Clearly the object has changed. It should now be apparent that even if we had detected the same attributes each time, it is still possible that the object in the second observation is not the same object as in the first observation. Even if we assume objects have some sort of unique identity independently of their state variables, we cannot know whether or not the object's unique identity has changed even if its state variables are unchanged. {\it Therefore we cannot track this supposed unique identity from one measurement to the next.} 

The distinguishing features of a particle (or any other system) {\em must lie purely in its observed state variables}. The only meaning of ``identity'' in quantum mechanics, is the object's {\em type}. Quantum mechanical particles have no observable unique existence except for the duration of an individual observation --- and then that uniqueness is contained purely in the state variables.

As an alternative argument, suppose that the concept of identity independent of the state variables was valid. Either it is detectable or it is not. If it is detectable, then we can simply include it in the state variables. If it is not detectable, then we can safely ignore it. The meaning of the wave function for a given system is then {\em not that it describes ``the'' object in a particular state, but that it describes ``an'' object in a particular state}.

\subsection{Identity And Indistinguishability}\label{sec:IdandInd}

Now that we have clarified the meaning of {\em identity} to refer to particle type, we also need to point out that, as we shall use the term throughout this paper, state {\em indistinguishability} is not the same thing. Two states that have the same identity may still be distinguishable by their variable quantum numbers (e.g. linear or angular momentum) even if those which define the identity (rest mass, charge, etc) are fixed. However, even in the case of all variable quantum numbers also being the same, two particle states may still be distinguishable if the other variables involved in their complete state descriptions are different (such as, for instance, a difference relating to their frame of reference). Hence we shall use the term indistinguishability in a much more general way to refer to any or all variables in a complete state description, and not just to the particle identity or its quantum numbers.

Throughout the rest of this paper, the reader should be aware, therefore, that it is the {\it complete} set of variables that compose the state description, not just the quantum numbers, that is essential for defining a single-valued state vector (see section \ref{sec:unique} for more explanation). Particle identity, of itself, is of no significance except that if two particles are indistinguishable then they must also be identical, although the converse is not true because identical particles may still be distinguishable by their variable quantum numbers or by other features of the way their states are described --- such as the way the frame of reference is chosen.

By the same token, none of what follows is specific to states of individual particles. Everything in this paper is equally true for composite states (and permutations of pairs of composite states) which have no specific mass shell or identity. In determining the permutation or exchange phase, and the physical consequences of permutation invariance, we concern ourselves only with the explicit state variables for each entity and any additional implicit distinguishability inherent in making those state descriptions complete. Hence the exclusion rules apply also to pairs of such composite states and the identities of any individual constituent particles are irrelevant, except where they are an explicit part of the state descriptions. (Composite states may be states of arbitrary constituents or specific constituents --- it doesn't matter which.) 

From now on, therefore, our discussion will center around the state variables and distinguishability necessary to uniquely describe the state. Any mention of ``particles'' rather than ``states'', or ``identity'' rather than ``distinguishability'', is essentially incidental and of no critical importance.\footnote{As another way to see the essential irrelevance of particle identity, note that, by allowing our space of states to cover the full range of all possible physical states and not just those of individual known particle types, we could claim that all such states are simply differing states of a single universal entity and hence that they all have the same ``id-entity''.}

\subsection{Permutation And Exchange}
We make a distinction between full permutation of {\it all} variables that form part of the {\it complete} state description necessary for uniqueness of the state vector, and the partial permutation, which we shall call ``exchange'', which takes place when any variables are specified in an order dependent way and therefore not properly permuted. In this latter case, exchange becomes equivalent to full permutation only when the unexchanged variables are indistinguishable.

By studying the general case of state permutation, whether distinguishable or not, and whether identical or not, and the full set of physical characteristics that distinguish states, we shall obtain some rules that enable us to understand state exchange for all pairs of physical states, whether they are distinguishable or not, and to discover the symmetry properties therein.

Although this connection and these distinctions are not usually expressed in this way, by doing so, and using an appropriately general and unambiguous notation, we shall refute the widely-held misconception that the exchange symmetry of state vectors for identical particles cannot be further determined without recourse to additional assumptions about nature, such as relativistic quantum field theory. 

No such additional assumption is necessary in classical physics --- where permutation invariance {\it by itself} reduces the number of possible independent states when tossing two identical coins from four permutations when the coins are distinguishable (e.g. toss one first and then the other) to three combinations when the coins are {\it in}distinguishable (e.g. toss both together and lose track of which is which). {\it It is our contention that, in quantum mechanics, just as in classical mechanics, the reduction in the number of states of two identical entities can be computed simply by recognizing any distinguishability in the system (including any order dependence in our state descriptions) and properly accounting for it --- thereby removing redundant permutations}.

\subsection{Permutation Invariance}\label{sec:perminv}
This section will summarize the notion of permutation invariance for multi-entity states whether distinguishable or not, and whether quantum mechanical or not. We define our permutation invariance assumption as:

\newtheorem{axiom}{Axiom}
\begin{axiom}[Permutation Invariance]\label{ax:perminv}
The physical properties of multiple entity states are independent of the order in which we observe or describe the collection of individual entities that make up the whole state.
\end{axiom}

The reader may consider this to be obvious and hardly worthy of explicit statement. However, we have chosen to make it explicit because some descriptions of exchange or permutation in quantum mechanics (e.g. the conventional and unqualified symmetrization postulate) actually violate this principle, as will become evident.

This permutation assumption applies equally to classical physics as to quantum physics. It also applies equally to distinguishable entities as to indistinguishable entities. It can be expressed mathematically in terms of individual state descriptions and collections of such state descriptions that do not necessarily have any relation to quantum mechanical state vectors.

Suppose we have several possibly distinguishable entities ``labeled'' (for want of a better word) $i,j,k...$. These could be individual particles, or more complex entities.  Suppose that their physical states are described by $S^i_a,S^j_b,S^k_c...$. These descriptions state that the entity distinguished by label $i$ is in a state $S^i_a$ and the entity distinguished by label $j$ is in a state $S^j_b$ and so on. However, as we have already argued, the distinguishing labels must correspond to physically significant features of the entity state if they are to provide observable distinctions. They may relate to entity type or method of description (e.g. the order in which coins are tossed, or the way frames of reference are defined) or some other feature. But if the entities are not distinguishable by the features that define the label, then the labels must be equal in value.

We do not, at this stage, nor the whole of this section and most of section \ref{sec:PA}, have to know anything about how these entities (or their states) are actually described (their state variables, etc), just that they {\it can} be described and that they may have different distinguishing labels, and, for section \ref{sec:PA}, that they have corresponding quantum mechanical state vectors in Hilbert space. Suppose also that these individual state descriptions are independent of each other and of the order in which we describe the entity states. The combined state is then described by a collection of individual states, which we can write as a list:
\begin{equation}
S^{ijk...}_{abc...} = S^i_a;S^j_b;S^k_c;... \label{eqn:collectdef}
\end{equation}

It is in the nature of lists that they are {\it ordered}. Although this collection is written as such a list, our permutation invariance assumption is that the properties of the collection are independent of the listing order. Hence all lists related by permuting the order in which the individual states appear are equivalent:
\begin{equation}
S_{abc...}^{ijk...} \equiv S_{bac...}^{jik...} \equiv etc \label{eqn:perminv}
\end{equation}
{\em and any one such list can stand in for any other as a description of the complete state}. In other words, the number of independent collections is given by the number of {\it combinations} of entity states rather than the number of permutations. 

Surprising though it may seem to those of us who are used to the conventional symmetrization postulate,  eqn. \ref{eqn:perminv} is sufficient, by itself, to tell us that any single-valued function $\psi$ of such multiple entity state descriptions must also be permutation symmetric:
\begin{equation}
\psi(S_{abc...}^{ijk...}) = \psi(S_{bac...}^{jik...}) = etc \label{eqn:funcperminv}
\end{equation}
Thus quantum-mechanical permutation can be related to classical permutation merely by finding a method to define single-valued wave functions for uniquely defined states.

In the case that two entities are indistinguishable by their labels ($i=j$). Then we find the additional property that
\begin{equation}
S_{abc...}^{iik...} \equiv S_{bac...}^{iik...} \equiv etc \label{eqn:redundantsuper}
\end{equation}

(Note that, in this case, permutation is equivalent to ``exchanging'' the state descriptions of the indistinguishable entities. We shall go into this in more detail in section \ref{sec:PA}.)

An important consequence of this indistinguishable entity symmetry (eqn. \ref{eqn:redundantsuper}) is a reduction in the number of independent collections. This is well-known in the case of coin-tossing to reduce the number of distinguishable combinations of two indistinguishable coins from four to three. In quantum mechanics it is known that this symmetry is connected to the identical particle exclusion rules but it is widely believed that these rules cannot be found from permutation invariance alone. Our purpose is to show that these exclusion rules can indeed be determined purely from permutation invariance as in the classical case --- as long as all physical distinguishability is properly accounted for.

\subsection{Order-Free Notation}\label{sec:orderfree}
It should be pointed out, however, that even with distinguishable entities, eqn. \ref {eqn:perminv} implies that we still have a redundancy in our notation. If we had a notation that had no order dependence in it, we could remove this redundancy. 

We could, for instance, write the individual states over the top of each other to illustrate the absence of any significance to the order in which we describe the individual entities. However, with normal two-dimensional media, such as paper, this is likely to result in illegibility.

One notation that could remove this redundancy would be to use a table with a column for each possible state of each type of distinguishable entity, in a fixed order (e.g. ascending order for numeric values) and in which the table entries simply specified the number of entities present in the state identified by the column index (a single entity state description). For example, a state of two indistinguishable tossed coins could be represented by a table with two columns labeled $H$ and $T$ (one for each allowed state). The entry in each column would be the number of coins in that state. Hence, instead of listing four states in the ordered notation: $(H;H)$, $(H;T)$, $(T;H)$ and $(T;T)$, we would find that there are only three distinguishable states describable in the unordered notation: $(2;0)$, $(1;1)$, and $(0;2)$. In general, states would be specified purely by the population numbers of the distinguishable allowed states.

The reason for drawing the readers attention to this alternative order-free notation is to point out that in principle it is just as possible to use such a notation to describe quantum-mechanical states as it is to describe classical states such as tossed coins. To avoid potentially infinitely large tables when continuous quantum numbers are considered, we could simply omit all states that have zero populations, and attach a state description to the columns that remain.

This gives us a method of constructing order-free state vectors and therefore leads to the important conclusions that (1) {\it any order dependence in quantum-mechanical state vectors is a consequence of an order dependent method of constructing the state vectors} --- whether dealing with distinguishable particles or indistinguishable particles, and (2) {\it The reduction in the number of states for two or more indistinguishable entities is a consequence of permutation invariance alone and requires no other assumptions}. In the case that the state labels are indistinguishable ($i=j$), permutation is equivalent to exchange and, if the state vectors are constructed in an order-free way, then the permutation and exchange eigenvalues will be $+1$.

It remains to show how order dependence arises in conventional ways of describing particle states and how such constructions enable us to derive the exclusion rules.

\section{Phase Ambiguity In State Vectors}
\label{sec:PA}

A common source of confusion when discussing the spin-statistics theorem is the notion that there is an arbitrary phase multiplier for any state vector. This often leads to the supposition that the exchange phase relating two identical particle state vectors which differ only in the ordering of the individual particles cannot be uniquely determined without some additional assumption.

The argument goes that particle exchange is a new discrete operation $X$ which can change the phase of the state vector. Since a repeat exchange recovers the original state vector ($X$ is its own inverse), then the eigenvalues of $X$ are $\pm 1$.

Our claim here, however, is that unless exchange of indistinguishable particles has some physical significance, then it is nothing more than permutation of the individual particle descriptions in the state vector which {\it by itself} can be of no significance in a state description. 

To see this, let us define a permutation operator $P$ such that:
\begin{equation}\label{eqn:permop}
P\ |S_{ab}^{ij}> = |S_{ba}^{ji}>
\end{equation}
whereas the exchange operator $X$ is such that:
\begin{equation}\label{eqn:xchgop}
X\ |S_{ab}^{ij}> = |S_{ba}^{ij}>
\end{equation}

Clearly, in the case $i = j$, we find $X$ and $P$ have the identical effect.

We have also shown that it is possible to define order-free state descriptions. Since these can also have quantum mechanical state vectors assigned to them, it is clearly possible to define order-free state vectors for which permutation is therefore the {\it identity} operation and hence that the eigenvalue of $P$ is always $+1$ and, therefore, in the case of indistinguishability, the eigenvalue of $X$ is also $+1$. Hence the listing order can be relevant only when it is linked to some order-dependence in the individual particle state descriptions and/or their notation and particle exchange $X$ can have an eigenvalue of $-1$ only when the particles are distinguishable in their labels ($i\ne j$). Some further explanation or qualification is clearly necessary. In particular we need to examine the {\it uniqueness} properties of state vectors and look closely at the significance of phase ambiguity.

It should also be plain from eqns. \ref{eqn:permop} and \ref{eqn:xchgop} that the nature of the exchange operator and its difference from plain permutation, is not some mystery that requires further theory to explain, but is purely an artifice of the superscripts $i,j$. From the discussion so far, it should be clear that these are simply a means of expressing that part of the individual state descriptions we choose to separate out so that they are not permuted with the rest of the state variables when we apply the exchange operator. These could be particle types only, in which case, exchange for identical particles is the same as permutation, or, as we shall shortly see, they could also include some other feature, such as an order-dependent definition of the spin quantization frame for each particle.

\subsection {Uniqueness Of State Vectors For Physically Complete State Descriptions}\label{sec:unique}
Although it is true we can choose our state vector for any given state from an infinite set of state vectors that differ only by a phase factor (given a specific normalization), we are always free to choose {\it one} such vector to be the unique representative of our physical state. Not only that, but we {\it must} choose such single-valued state vectors if we wish to calculate effects such as interference. Once we have a unique prescription for making that choice for any state description, then we no longer have any phase ambiguity unless we change the way we describe the state. Hence any residual phase ambiguity arises solely from any residual ambiguity in the state description that renders it insufficient to define a single-valued state vector. The proof of this is trivial. Suppose $S$ and $S'$ are two alternative descriptions of the same physical state. $S$ can be a single particle state or a list of such states. Then
\begin{equation}
|S'> = f(S',S)\ |S>
\end{equation}
where $f(S',S)$ is a phase factor. Clearly, if $|S>$ is single-valued, then 
\begin{equation}
f(S,S) = 1. \label{eqn:uniqueness}
\end{equation}
So the existence of a phase change depends on the distinction between the state {\it descriptions} $S'$ and $S$, even though they represent the same physical state. 

To understand the type of distinction in the state description that can create such a phase change, consider the angular momentum representation (i.e. when we have states of definite $j$ and $m$) when we rotate the frame of reference about the angular momentum quantization axis. Although the frame of reference (which is part of the state description) changes ($S\rightarrow S'$), the physical quantum numbers remain unchanged. However, if both  $|S>$ and $|S'>$ lie in the same Hilbert space,  yet the transformation leaves the physical observables unchanged, they may differ at most by a phase. In fact that phase is uniquely determined by the product of the angle of rotation and the third component of angular momentum $m$.

Now suppose we have two descriptions of the same state that are physically {\it equivalent} ($\bar S \equiv S$). By this we mean that not only are the quantum numbers identical, but that all other physical features of the state, such as the method used to specify the frame of reference are also physically equivalent. Since there is no transformation on the frame of reference or any other physical transformation and no other physically observable difference, there can be no transformation in Hilbert space to correspond to the change in description $S\rightarrow \bar S$. If both state vectors are single-valued, exist in the same Hilbert space and yet cannot be related by any transformation except the identity operation then they must be identical:
\begin{equation}
|\bar S> = |S> \label{eqn:equiv}
\end{equation}

As an alternative proof, suppose that there {\it was} a phase difference ($f(\bar S,S)\not= 1$). Then the relationship between $\bar S$ and $S$ would be as physically significant as any physical transformation that produced the same change in phase - which would violate our definition of $\bar S$ that there was no such physical significance. (As an example of this, consider a change in description which produces $f(\bar S,S)=e^{i\eta}$. This phase change is also produced by a rotation of the spin quantization frame by $-\eta/m$ where $m$ is the component of spin along the axis of rotation. Hence the change of description $S\rightarrow \bar S$ is equivalent to such a rotation and our transformation $S\rightarrow \bar S$ must be physically significant. This is true for a single-particle state or a multi-particle state and, in the latter case, whether or not the particles are identical.)

It might be argued that a rotation by $2\pi$, or any multiple of $2\pi$, is not physically significant since the frame of reference is unchanged. Yet such a rotation can change the sign of a fermionic state vector. We would claim, however, that such a rotation is still a physically recognizable transformation\footnote{According to Feynman\cite{Feynman}, a ``rotation one time around can be distinguished from doing nothing at all'' and he cites Penrose and Ridler\cite{PenroseRidler} as giving an example from Dirac demonstrating this with a twisted arm holding a cup. The asymmetry central to this paper, distinguishing a clockwise $\pi$ rotation of one vector into another, from a counter-clockwise $\pi$ rotation, is an example, more directly pertinent to particle geometry, of this same effect.} on the frame of reference, even if the resulting frame of reference is recognizably different only in the context of how we got to it, and is therefore a physical transformation that can change our state description and therefore potentially change the phase of the state vector. In any case, it must be obvious that we must distinguish the two state descriptions before and after such a rotation if we wish our state vector to be single-valued.

The consequence of this is that we can always choose our state vectors so that any residual relative phase between two state vectors for the same state can be limited to situations where there is a physical difference between the ways we observe and describe the states (i.e. where this difference can be described by a transformation in Hilbert space).

We shall summarize this discussion in the form of a generalized physical completeness principle:

\begin{axiom}[Physical Completeness Principle]\label{ax:physuniq}
For any physically complete state description, it is possible to choose a unique state vector that is unchanged by any non-physical transformation.
\end{axiom}

Clearly this principle rests on the notion of what makes a state description ``physically complete''. From the above discussion, it must be plain that the difference between a ``physically complete state description'' and a complete set of quantum numbers for a given physical state lies in the fact that the former includes the elimination of any physical transformations that leave the quantum numbers unchanged (including the elimination of arbitrary rotations of the spin quantization frame about the quantization axis or by $2\pi$ about any arbitrary axis). We shall show how to eliminate such arbitrary rotations in the case of a single particle or a pair of particles in sections \ref{sec:USQ} and \ref{sec:spatasym}. For the time being, we shall take it as self-evident that any alternative ``physically complete'' descriptions describing the same state that are related only by non-physical transformations are {\em physically equivalent} in the sense discussed above and therefore, from axiom \ref{ax:physuniq}, satisfy eqn. \ref{eqn:equiv}. In other words, {\em physically indistinguishable state descriptions have identical state vectors}. Although a unique set of quantum numbers is insufficient to specify a unique state vector, a physically complete description is sufficient to eliminate this ambiguity.

State descriptions that are related by a physical transformation, however, have state vectors that are related by an equivalent transformation in Hilbert space. Even when the physical transformation leaves the quantum numbers unchanged, then  the state descriptions are not physically equivalent, since the transformation describes the relationship between the additional features which distinguish the state descriptions (such as might be implied by the labels introduced in the previous section). The Hilbert space transformation then defines the phase change that relates the different state vectors. 

\subsection{A Revised Symmetrization Postulate}\label{sec:revsympos}
We argued in the last section that unless the state descriptions for two state vectors describing the same physical state differ by a physically significant transformation, then we can always choose these two state vectors to be identical. In particular, unless you consider particle permutation to be a physically significant transformation, (contrary to our permutation invariance assumption) then uniqueness implies that pure permutation cannot, {\it by itself}, introduce any  change of phase between multi-particle state vectors. We shall express this in a revised symmetrization postulate: 

\newtheorem{theorem}{Theorem}
\begin{theorem}[Revised Symmetrization Postulate]\label{th:symmpost}
State vectors for multiple entity states that are described in a physically complete and order independent way can be chosen to be symmetric under permutation.
\end{theorem}

The proof of this theorem follows trivially from axioms \ref{ax:perminv} and \ref{ax:physuniq}. In particular, it is true, {\em regardless of particle spin}.

As a consequence, any ``exchange (or permutation) asymmetry'' can arise only through a physically significant order dependent asymmetry (whether explicit or implicit) in the individual particle state descriptions.

As further clarification, let us explain the connection between order dependent and order independent state vectors.

Suppose we have two potentially order-dependent state vectors describing the same physical state. They can differ, at most, by a phase factor:
\begin{equation}\label{eqn:permasymstatevect}
|S^j_b;S^i_a> = f((S^j_b;S^i_a)\leftarrow (S^i_a;S^j_b))\ |S^i_a;S^j_b>
\end{equation}

Suppose we now use an order free notation to describe the state. Since the state descriptions are order independent, then, if they are also physically complete,  this state vector clearly satisfies our revised symmetrization postulate. As discussed in secs. \ref{sec:perminv} and \ref{sec:orderfree}, we shall write the state vector as 
\begin{eqnarray}\label{eqn:orderfreestatevect}
|S^{ij}_{ab}> \ = \ |S^{ji}_{ba}> & = & |S^i_a (1);S^j_b(1)>\nonumber\\
|S^{ii}_{aa}> & = & |S^i_a (2)>
\end{eqnarray}
where $S^i_a < S^j_b$ by some arbitrary criteria and the digits in parentheses are the population numbers of the states. 

Clearly the state vectors in eqns. \ref{eqn:orderfreestatevect} are related to the order dependent state vectors by again, at most, a phase factor:
\begin{equation}\label{eqn:perminvstatevect}
|S^{ij}_{ab}> =  f(S^i_a;S^j_b)\ |S^i_a;S^j_b> = f(S^j_b;S^i_a)\ |S^j_b;S^i_a>
\end{equation}
and the permutation phase factor in eqn. \ref{eqn:permasymstatevect} is determined by:
\begin{equation}
f((S^j_b;S^i_a)\leftarrow (S^i_a;S^j_b)) = {f(S^i_a;S^j_b)\over f(S^j_b;S^i_a)}
\end{equation}

For the rest of this paper, we shall always use order independent state descriptions (except where the order dependence is implicit in the labels $i,j$, rather than the listing order). Hence, we shall always have:
\begin{equation}\label{eqn:ordind}
f(S^i_a;S^j_b) = f(S^j_b;S^i_a) = 1
\end{equation}

\subsection{The Origin Of An Exchange Phase}\label{sec:orgnxphase}
If our listing-order is purely a matter of notation and has no significance for the description of the individual states, then we have seen that uniqueness requires that we can define state vectors that are order independent and permutation symmetric, and, in the case of indistinguishable particles, this permutation symmetry alone will be a filter for the permitted states. We shall now show how an exchange phase can nevertheless arise in situations where we use a notation in which the particle ordering affects the individual descriptions of the individual particles. Then the exchange phase of the state vectors will be determined by the Hilbert space transformations brought about by {\it changes in the individual descriptions} resulting from the change in ordering.

In general, as shown explicitly in the last section (eqn. \ref{eqn:ordind}), permutation invariance and uniqueness of state vectors for equivalent state descriptions imply that we can always define permutation symmetric state vectors:
\begin{equation}
|S^i_a;S^j_b;...> =  |S^{ij...}_{ab...}> = |S^{ji...}_{ba...}> = |S^j_b;S^i_a;...>\label{eqn:orderindep}
\end{equation}

Now, we saw in subsection \ref{sec:unique} that state vectors for distinguishable particles are related by the transformations in Hilbert space corresponding to the physical transformations that relate their distinguishing state descriptions. Hence, even for states of identical particles with identical quantum numbers, this distinguishability implies that
\begin{eqnarray}
S_a^i \not\equiv S_a^j\\
S_b^i \not\equiv S_b^j
\end{eqnarray}

For such states, where the distinguishability results purely from a difference in description for what is ostensibly the same physical state, then we can define both state vectors in the same Hilbert space and differing by, at most, a phase:
\begin{eqnarray}
|S_a^i> & = & f(S_a^i\leftarrow S_a^j)\ |S_a^j>\nonumber\\
|S_b^i> & = & f(S_b^i\leftarrow S_b^j)\ |S_b^j>\label{eqn:singlephase}\\
|S_{ab...}^{ij...}> & = & f(S_{ab...}^{ij...}\leftarrow S_{ba...}^{ij...})\ |S_{ba...}^{ij...}>\nonumber\\
|S_{ab...}^{ij...}> & = & f(S_{ab...}^{ij...}\leftarrow S_{ab...}^{ji...})\ |S_{ab...}^{ji...}> \label{eqn:doublephase}
\end{eqnarray}
and we find, from eqn. \ref{eqn:orderindep}, that
\begin{eqnarray}
f(S_{ab...}^{ij...}\leftarrow S_{ba...}^{ij...}) & = &  f(S_{ab...}^{ij...}\leftarrow S_{ab...}^{ji...})\\
f(S_{ab...}^{ij...}\leftarrow S_{ba...}^{ji...}) & = & 1
\end{eqnarray}
{\it In other words, any exchange phase that might arise is a consequence not of the particle permutation but of the exchange of distinguishing characteristics ($i\leftrightarrow j$).}

In the next section we address the question of the relation between the single-particle distinguishability phase factors and the two-particle exchange phase factors.

\subsection{Determination Of The Exchange Phase}\label{sec:detxphase}
To relate the multi-particle exchange phases to single-particle distinguishability phases we need to know how to relate transformations in multi-particle space to those in single-particle space. We do this by relating the multi-particle state vectors to the single-particle state vectors. 

Multi-particle state vectors can be chosen as direct product state vectors from the Hilbert space that arises from the direct product of the single-particle Hilbert spaces. Assuming a different Hilbert space $H^i,H^j,...$ etc. for each distinguishable entity $i,j,...$, then for distinguishable entities ($i\ne j$), always described in a particular order ($i$ first), the state vector will lie in the direct product space $H^i\otimes H^j$. If described in reverse order ($j$ first) then it would lie in $H^j\otimes H^i$. These two different product vectors lie in two distinct spaces. To define a permutation invariant and order-independent state vector obeying eqn. \ref{eqn:orderindep} we take the symmetrized linear combination of the direct product vectors:
\begin{equation}
|S_{ab}^{ij}> = \alpha (|S_a^i>|S_b^j>\ +\ |S_b^j>|S_a^i>) \label{eqn:directprod1}
\end{equation}
where the factor $\alpha$ is for normalization only. 

An alternative way to look at this would be to absorb the labels $i,j,...$ into the state descriptions in a single Hilbert space $H$ for all particles. Then the composite space is the direct product space $H\otimes H$. Clearly, all physical states must lie in the symmetric subspace only, if the state vectors are to be single-valued and permutation symmetric and again we find we can use eqn. \ref{eqn:directprod1}.

We also have
\begin{equation}
|S_{ba}^{ij}> = \alpha (|S_b^i>|S_a^j>\ +\ |S_a^j>|S_b^i>) \label{eqn:directprod2}
\end{equation}
When $S^i_a$ and $S^j_a$, have the same quantum numbers but are distinguishable by the labels $i,j$, and similarly for $S^i_b$ and $S^j_b$, substituting the single particle distinguishing phases of eqn. \ref{eqn:singlephase} in eqn. \ref{eqn:directprod1} enables us to compute the two-particle exchange phase in eqn. \ref{eqn:doublephase}:
\begin{equation}
f(S_{ab...}^{ij...}\leftarrow S_{ba...}^{ij...}) = {f(S_a^i\leftarrow S_a^j) \over f(S_b^i\leftarrow S_b^j)}
\end{equation}

{\it Thus the exchange phase is computed purely from the single particle phase changes that arise from the exchange of distinguishing features $i$ and $j$, and we have indicated how it can be done in the general case without introducing any special additional assumptions such as relativity or local field theory.}

However, we would stress that the only reason for introducing these exchange phase factors is when there is a genuine physical asymmetry (which implies distinguishability) in the way the individual particle states are described, corresponding to a transformation in Hilbert space when the distinguishing features are exchanged. Without exchanging such distinguishing features of the individual state descriptions, uniqueness requires that the phase factor obtained by simple re-ordering would always be unity.

\subsection{Uniqueness And Spin Quantization}\label{sec:USQ}
In this subsection we review the definition of a state vector for a single particle of arbitrary spin and show how to define a physically complete state description and a corresponding single-valued state vector in the sense outlined in section \ref{sec:unique}.

Conventionally, a state vector of arbitrary momentum $\mathbf{p}$ and spin component $m$ along an axis $\hat{\mathbf{n}}$ is defined by\cite{Wigner}:
\begin{equation}\label{eqn:wigdef}
|Q,\mathbf{p},s,m(\hat{\mathbf{n}})> = U(B(\mathbf{p}))\ |Q,\mathbf{0},s,m(\hat{\mathbf{n}})>
\end{equation}
where $|Q,\mathbf{0},s,m(\hat{\mathbf{n}})>$ is a rest frame eigenstate of $\mathbf{J}^2$ and component $J_{\hat{\mathbf{n}}}$ (in the direction $\hat{\mathbf{n}}$) with eigenvalues $s(s+1)$ and $m$, $U(B(\mathbf{p}))$ is the operator describing the boost which takes the momentum from $\mathbf{0}$ to $\mathbf{p}$:
\begin{equation}\label{eqn:boostdef}
U(B(\mathbf{p}))\ = U(R(\hat{\mathbf{z}}\rightarrow\hat{\mathbf{p}}))U(B(p\hat{\mathbf{z}}))U(R^{-1}(\hat{\mathbf{z}}\rightarrow\hat{\mathbf{p}}))
\end{equation}
and $Q$ represents all other intrinsic quantum numbers. We remind the reader, in passing, that the rotation $R(\hat{\mathbf{z}}\rightarrow\hat{\mathbf{p}})$ is defined to be that which takes the z-axis of the frame of reference from the direction of motion $\hat{\mathbf{p}}$ into the z-axis of the {\it final} frame of reference, therefore transforming the momentum from $p\hat{\mathbf{z}}$ to $\mathbf{p}$.

We note that in Wigner's treatment, $B(p\hat{\mathbf{z}})$ is a Lorentz boost. However, it is worth noting here that in what follows, this not need be the case, but it could equally well be a Galilean boost. If we had  chosen to use co-ordinate space rather than momentum space, then we could also replace the boost by a translation. It should soon become apparent that what matters for us can be contained purely in the spin part of the state vector, but we give a fuller description in momentum space simply to provide a more complete setting.

Two common choices of spin quantization axis $\hat{\mathbf{n}}$ are:
\begin{enumerate}
\item The {\it canonical} basis ($\hat{\mathbf{n}} = \hat{\mathbf{z}}$) in which the spin quantization axis is the z-axis of the frame in which the momentum is measured.
\item The {\it helicity} basis ($\hat{\mathbf{n}} = \hat{\mathbf{p}}$) in which the spin quantization axis is parallel to the momentum.
\end{enumerate} 

Unfortunately, it isn't hard to see that both of these choices of state vector are potentially ambiguous up to an arbitrary rotation about their spin quantization axis, because none of the explicit state variables are changed by such a rotation. To obtain a single-valued state vector, we must look at what happens under rotations of the frame of reference.

Under a rotation $R(\hat{\mathbf{n}}'\rightarrow \hat{\mathbf{n}})$ the rest frame vector transforms as (e.g. \cite{Rose}) 
\begin{eqnarray}
|Q,\mathbf{0},s,m(\hat{\mathbf{n}})> & = & U(R(\hat{\mathbf{n}}'\rightarrow \hat{\mathbf{n}}))\ |Q,\mathbf{0},s,m(\hat{\mathbf{n}}')>\nonumber\\
& = & \sum_{m'}\ D^s_{m'm}(R(\hat{\mathbf{n}}'\rightarrow \hat{\mathbf{n}}))\ |Q,\mathbf{0},s,m'(\hat{\mathbf{n}}')>
\end{eqnarray}
and therefore, for general $\mathbf{p}$, state vectors with differing spin quantization frames are related by the rotation which relates those frames:
\begin{eqnarray}\label{eqn:rotatespinframe}
|Q,\mathbf{p},s,m(\hat{\mathbf{n}})>
& = & \sum_{m'}\ D^s_{m'm}(R(\hat{\mathbf{n}}'\rightarrow \hat{\mathbf{n}}))\ |Q,\mathbf{p},s,m'(\hat{\mathbf{n}}')>
\end{eqnarray}

As long as the rotation does not change the spin quantization axis, then of course, the effect of the rotation is at most a change in phase, determined by the angle of rotation $\gamma$ about $\hat{\mathbf{n}}$:
\begin{equation}
D^s_{m'm}(R_{\hat{\mathbf{n}}}(\gamma)) = \delta_{m'm}e^{-im\gamma}
\end{equation}

Substituting into eqn. \ref{eqn:rotatespinframe}, with $\hat{\mathbf{n}}'=\hat{\mathbf{n}}$, it is easy to see that the notation of eqn. \ref{eqn:wigdef} is not sufficient for single-valued state vectors. For integer spin it is sufficient to specify the complete quantization frame rather than just $\hat{\mathbf{n}}$ to eliminate this ambiguity, but for half-integer spin, even this is insufficient. {\it To eliminate the ambiguity completely for general spin we must first of all specify a standard base frame that is implicit in all state vectors, then specify the rotation which takes this base frame into the spin quantization frame}.

When dealing with single particle state vectors, or non-identical particles, it is usually obvious that the implied base frame is the same frame of reference in which we measure $\mathbf{p}$. The remaining $2\pi$ ambiguity doesn't concern us much because we don't usually anticipate any $2\pi$ rotations taking place when defining $\mathbf{p}$ in that frame. However, as we have seen, the derivation of an exchange phase is crucially dependent on resolving any such ambiguity in order to obtain single-valued state vectors; so we must be careful to make the state variables which can change the phase explicit. For the single particle state, we therefore define, instead of eqn. \ref{eqn:wigdef},
\begin{equation}\label{eqn:newwigdef}
|Q,\mathbf{p},s,m(R_{BS})>_B = U(B(\mathbf{p}))\ |Q,\mathbf{0},s,m(R_{BS})>_B
\end{equation}
where the suffix $B$ indicates the choice of standard base frame and $R_{BS}$ is the rotation which takes the base frame into the spin quantization frame $S$. When $B$ is the same (canonical) frame of reference $C$ in which we measure $\mathbf{p}$, we have
\begin{equation}
R_{BS} = R_{CS} = R(\hat{\mathbf{z}}\rightarrow \hat{\mathbf{n}})
\end{equation}

Whenever $R_{BS}$ is the null rotation $N$ (the spin quantization frame is also the base frame), we shall omit it and define:
\begin{eqnarray}\label{eqn:nulldef}
|Q,\mathbf{p},s,m>^B = |Q,\mathbf{p},s,m(N)>_B
\end{eqnarray}
and we note, in passing that the same coincidence of spin quantization frame and base frame is also true for
\begin{eqnarray}
|Q,\mathbf{p},s,m(R_{\hat{\mathbf{q}}}(2\pi))>_B = (-1)^{2s} |Q,\mathbf{p},s,m>^B
\end{eqnarray}
where $R_{\hat{\mathbf{q}}}(2\pi)$ is a rotation by $2\pi$ about any arbitrary axis $\hat{\mathbf{q}}$. However, by using the notation of eqn. \ref{eqn:newwigdef} and the definition of eqn. \ref{eqn:nulldef} we can distinguish these cases and avoid the sign ambiguity.

When the spin quantization frame is also the canonical frame ($S=C$), 
\begin{eqnarray}\label{eqn:canondef}
|Q,\mathbf{p},s,m>^C = |Q,\mathbf{p},s,m(N))>_C
\end{eqnarray}

Similarly, in the helicity basis, we can define 
\begin{eqnarray}
|Q,\mathbf{p},s,\lambda>^H = |Q,\mathbf{p},s,\lambda(N)>_H
= |Q,\mathbf{p},s,\lambda(R(\hat{\mathbf{z}}\rightarrow \hat{\mathbf{p}}))>_C \end{eqnarray}

The relationship between the canonical and helicity state vectors is then given by eqn. \ref{eqn:rotatespinframe}:
\begin{eqnarray}\label{eqn:heltocan}
|Q,\mathbf{p},s,m>^C
& = & \sum_{\lambda}\ D^s_{\lambda m}(R(\hat{\mathbf{p}}\rightarrow \hat{\mathbf{z}}))\ |Q,\mathbf{p},s,\lambda>^H
\end{eqnarray}

Under an arbitrary rotation R, which transforms the frame of reference in which the momentum is $\mathbf{p}$ into one in which the momentum is $\mathbf{p}'$, the general state vector defined in eqn. \ref{eqn:newwigdef} transforms according to:
\begin{eqnarray}
\lefteqn{U(R)|Q,\mathbf{p},s,m(R_{BS})>_B}\nonumber\\ & = & U(B(\mathbf{p}'))U(R)|Q,\mathbf{0},m(R_{BS})>_B\nonumber\\
& = & U(B(\mathbf{p}'))|Q,\mathbf{0},m(R.R_{BS})>_B\nonumber\\
& = & |Q,\mathbf{p}',s,m(R.R_{BS})>_B\nonumber\\
& = & \sum_{m'}\ D^s_{m'm}(R)\ |Q,\mathbf{p'},s,m'(R_{BS})>_B
\end{eqnarray}
and 
\begin{eqnarray}
R_{BS'}=R(\hat{\mathbf{p}}\rightarrow \hat{\mathbf{p}}').R_{BS}=R.R_{BS}
\end{eqnarray}
is the rotation which takes the base frame into the rotated spin-quantization axis in the rotated system.

In the canonical basis:
\begin{eqnarray}
U(R)|Q,\mathbf{p},s,m>^C & = & U(R)|Q,\mathbf{p},s,m(N)>_C\nonumber\\
& = & \sum_{m'}\ D^s_{m'm}(R)\ |Q,\mathbf{p}',s,m'(N)>_C\nonumber\\
& = & \sum_{m'}\ D^s_{m'm}(R)\ |Q,\mathbf{p}',s,m'>^C
\end{eqnarray}
and we see that the third component of spin is transformed by the rotation.

In the helicity basis, however, the helicity frame transforms with the momentum:
\begin{eqnarray}
U(R) |Q,\mathbf{p},s,\lambda>^H = U(R)|Q,\mathbf{p},s,m(R(\hat{\mathbf{z}}\rightarrow \hat{\mathbf{p}}))>_C\nonumber\\
= |Q,\mathbf{p}',s,(R(\hat{\mathbf{z}}\rightarrow \hat{\mathbf{p}}'))>_C = |Q,\mathbf{p}',s,\lambda>^H
\end{eqnarray}
and the helicity is therefore unchanged by the rotation.

Now the question arises of how to generalize this notation and the specification of a standard base frame to the case of two particles. It might na\"ively be thought that we could simply choose the same (canonical) base frame for both particles. Indeed, this is implicit in the conventional approach. However, as we shall see in the next section, it is not possible to do this without introducing a new asymmetry between the particles. To be safe, it is better to define independent base frames for each particle in a symmetric way. This ensures that we can independently define physically complete descriptions for each particle and it will be seen that this will enable us to completely understand the origin of the conventional exchange antisymmetry of identical fermions.

\section{Geometrical Asymmetry For Particle Pairs} \label{sec:spatasym}
We have previously and frequently alluded to an inherent spatial asymmetry in two-particle states. In this section we shall provide a detailed discussion of this asymmetry and its consequences for defining physically complete state descriptions and unique state vectors.

\subsection{Asymmetry In A Common Frame Of Reference}\label{sec:asymcommon}
Our purpose here is to explain why it is not possible to choose a common frame of reference for two particles in a way that is symmetrical (does not distinguish) between the orientations of the individual particles. One can choose either a common frame of reference or distinct symmetrically defined frames of reference for each particle. But one cannot do both simultaneously. If this asymmetry (distinction) in a common frame is not properly accounted for in the state variables, then it must be specified in some other way (e.g. as a distinguishing label --- implying a corresponding exchange asymmetry) if we are to have physically complete state descriptions and uniquely defined state vectors (in the sense of unique relative phases discussed in section \ref{sec:PA}).

The state description of each particle in the system has at least one physical vector $\mathbf{v}$ attached to it. It could be the position vector $\mathbf {r}$ (in co-ordinate space), the linear momentum vector $\mathbf {p}$ (in linear momentum space), the spin quantization axis $\hat{\mathbf {n}}$ or any other physical vector that is part of the state description of that particle. In a common frame of reference, each vector is described with respect to the same frame.

In quantum mechanics, a single-valued state vector for a single particle requires the unique specification of the rotation $R(\hat{\mathbf{z}}\rightarrow \hat{\mathbf{v}})$ which takes the z-axis of the frame of reference into its physical vector $\mathbf{v}$ or {\it vice versa}. In the last section, for instance, we saw that we must uniquely specify the rotation which takes the z-axis into the spin quantization axis.

In a two-particle state, this has to be done for {\it both} particles. For unique, individual state vectors these rotations ($R_a=R(\hat{\mathbf{z}}\rightarrow \hat{\mathbf{v}}_a)$ and $R_b=R(\hat{\mathbf{z}}\rightarrow \hat{\mathbf{v}}_b)$) must be defined independently for each particle or the difference must be accounted for in the state descriptions. The question then arises: Is it possible to define these orientations in a way that is symmetric between both vectors and obtain the same frame of reference for both particles? And, if not, what are the consequences for the state vector in a common frame of reference?

\begin{figure}[htbp]\begin{center}
\caption{\label{fig:vvx} Impossibility Of A Symmetric Common Set Of Axes}
\setlength{\unitlength}{1in}
\begin{picture}(4,1.8)
\put(1,1){\vector(0,1){0.5}}
\put(0.95,1.6){$\hat{\mathbf{y}}_a$}
\put(1,1){\vector(3,1){1.0}}
\put(2.1,1.3){$\hat{\mathbf{v}}_b$}
\put(1.6,1.05){$\theta$}
\put(1,1){\vector(3,0){1.5}}
\put(2.6,0.98){$\hat{\mathbf{k}}=\hat{\mathbf{z}}$}
\put(1.8,0.85){$\theta$}
\put(1,1){\vector(3,-1){2.0}}
\put(3.1,0.3){$\hat{\mathbf{v}}_a$}
\put(1,1){\vector(0,-1){0.5}}
\put(0.95,0.3){$\hat{\mathbf{y}}_b$}
\end{picture}
\end{center}
\end{figure}
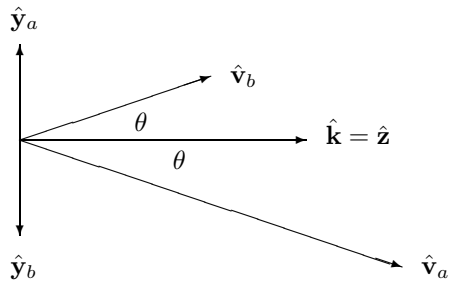

Let us start by choosing a symmetrically defined z-axis. This is easy to do. We choose the axis $\mathbf{k}$ which bisects the two vectors. Each vector will then make an angle
\begin{eqnarray}\label{eqn:deftheta}
\theta = {\cos^{-1}(\hat{\mathbf{v}}_a.\hat{\mathbf{v}}_b)\over 2}
\end{eqnarray}
with the common z-axis. (See fig. \ref{fig:vvx}.) It matters not whether we choose $\theta$ to be acute or obtuse {\it as long as we preserve symmetry by making the same choice for both particles}. 

Now let us see if we can choose a common y-axis symmetrically. Like the z-axis, this must lie in the plane which bisects $\hat{\mathbf {v}}_a$ and $\hat{\mathbf {v}}_b$. It must also be perpendicular to the z-axis. Hence it must be given by either
\begin{equation}
\hat{\mathbf{y}} = \hat{\mathbf{v}}_a\times\hat{\mathbf{v}}_b
\end{equation}
or
\begin{equation}
\hat{\mathbf{y}} = \hat{\mathbf{v}}_b\times\hat{\mathbf{v}}_a
\end{equation}

Each choice of y-axis (and x-axis) is asymmetric between the particles. If we choose
\begin{equation}
\hat{\mathbf{y}}_a = \hat{\mathbf{v}}_a\times\hat{\mathbf{v}}_b
\end{equation}
then
\begin{equation}\label{eqn:symydef}
\hat{\mathbf{y}}_b = \hat{\mathbf{v}}_b\times\hat{\mathbf{v}}_a = -\hat{\mathbf{y}}_a
\end{equation}
and we can independently and symmetrically define $\hat{\mathbf{y}}_a$ and $\hat{\mathbf{y}}_b$, but, since they are opposite to each other, they cannot coincide. (See fig. \ref{fig:vvx})
Furthermore, the x-axis can then not be in the bisecting plane, and therefore favors one particle over the other.

This asymmetry ($\hat{\mathbf{y}}_b = -\hat{\mathbf{y}}_a$) persists even in the limit $\hat{\mathbf {v}}_a\rightarrow \hat{\mathbf {v}}_b$. Even with a common z-axis, we cannot also choose a common y-axis without preferring one particle over the other. Hence {\it we cannot choose a common frame of reference for both particles without introducing an asymmetry (and therefore a possible exchange phase) between the particles}.

\subsection{Accounting For The Asymmetry}\label{sec:asymaccount}
The existence of this asymmetry means that we cannot, simply by specifying a common frame, assume that state descriptions for two particles are both physically complete {\it and} order-independent, since the method of specifying the orientation of each particle in that common frame will differ and therefore imply an asymmetry. This asymmetry would potentially prevent computation of the relative phase under exchange. Since the whole point of this paper is to compute this relative phase for any given definition of the state vectors, we cannot permit such ambiguity in their definition. However, it is possible to account for this asymmetry and examine its effect on our state vectors, if we use the following prescription:

\begin{enumerate}
\item Define independent (but symmetrically prescribed) frames of reference for each particle. We know then that the frames of reference do not introduce any new asymmetry. Hence the individual state descriptions for each particle can be made physically complete in an order independent way and then, by our revised symmetrization postulate (theorem \ref{th:symmpost} in section \ref{sec:revsympos}), the two-particle state vector can then be defined to be permutation symmetric.
\item By examining the effect of the rotations which then take each particle's independent frame of reference into a common frame we can compute the effect of the asymmetry in the common frame.
\end{enumerate}

We shall now turn our attention to defining independent frames of reference in a way that is symmetrical between the particles. 

\subsection{Independent ``Parallel'' Frames}\label{sec:parallel}
For instance, using the subscript $c$ for the {\it current} particle and $o$ for the {\it other} particle, we can choose the z-axis for each particle to be parallel to its vector $\hat{\mathbf{v}}$:
\begin{eqnarray}\label{eqn:parallelz}
\hat{\mathbf{z}}_c = \hat{\mathbf{v}}_c
\end{eqnarray}
and the y-axes by their cross-product:
\begin{equation}\label{eqn:currenty}
\hat{\mathbf{y}}_c = \hat{\mathbf{v}}_c\times\hat{\mathbf{v}}_o
\end{equation}
which clearly gives us a symmetric method for choosing independent frames. We shall call this choice of y- and z-axes, for each particle, its {\it parallel} frame. 

\subsection{Independent ``Bisecting'' Frames}\label{sec:bisecting}
Alternatively, choosing a common z-axis as the axis $\hat{\mathbf{k}}$ which bisects the two vectors:
\begin{eqnarray}\label{eqn:bisectingz}
\hat{\mathbf{z}}_a = \hat{\mathbf{z}}_b = \hat{\mathbf{k}}
\end{eqnarray}
and the y-axes again by their cross-product (eqn. \ref{eqn:currenty}) gives us another symmetric method for choosing independent frames. We shall call this choice of y- and z-axes, for each particle, its {\it bisecting} frame. 

\subsection {General Symmetrically-Defined Independent Frames}\label{sec:genind}
The parallel and bisecting frames are related by the same rotation about the y-axis $R_{\hat{\mathbf{y}}}(\theta)$, for both particles. In general by applying the same arbitrary rotation to the parallel frames for both particles we can generate pairs of independent symmetrically-defined frames for any orientation of axes we like.

The relationship between the parallel frames of reference is a rotation of $\pm\pi$ about the axis $\hat{\mathbf{k}}$. This same rotation will also take one particle's vector into that of the other:
\begin{eqnarray}\label{eqn:asymrotparallel}
R(\mathbf{z}_a\rightarrow \mathbf{z}_b) & = R(\mathbf{v}_a\rightarrow \mathbf{v}_b) & = R_{ab} = R_{\mathbf{k}}(\pm\pi)\nonumber\\
R(\mathbf{z}_b\rightarrow \mathbf{z}_a) & = R(\mathbf{v}_b\rightarrow \mathbf{v}_a) & = R_{ba} = R_{ab}^{-1} = R_{\mathbf{k}}(\mp\pi)
\end{eqnarray}

A similar rotational relationship holds for the bisecting frames:
\begin{eqnarray}\label{eqn:asymrotbisect}
R(\mathbf{y}_a\rightarrow \mathbf{y}_b) & = R(\mathbf{v}_a\rightarrow \mathbf{v}_b) & = R_{ab} = R_{\mathbf{k}}(\pm\pi)\nonumber\\
R(\mathbf{y}_b\rightarrow \mathbf{y}_a) & = R(\mathbf{v}_b\rightarrow \mathbf{v}_a) & = R_{ba} = R_{ab}^{-1} = R_{\mathbf{k}}(\mp\pi)
\end{eqnarray}

In general, for any pair of symmetrically-defined independent frames the rotation which relates those frames is given by:
\begin{eqnarray}\label{eqn:asymrot}
R(\mathbf{v}_a\rightarrow \mathbf{v}_b) & = & R_{ab} = R_{\mathbf{k}}(\pm\pi)\nonumber\\
R(\mathbf{v}_b\rightarrow \mathbf{v}_a) & = & R_{ba} = R_{ab}^{-1} = R_{\mathbf{k}}(\mp\pi)
\end{eqnarray}

It doesn't matter whether we choose a clockwise rotation ($+\pi$) for $R_{ab}$ or an anti-clockwise rotation ($-\pi$): $R_{ba}$ will always be in the opposite direction.

Whatever pair of symmetrically-defined independent frames we choose, we may always select either frame as a common frame of reference as long as we explicitly account for the rotation of the independent frame of one particle into that of the other. But in doing so, because $R_{ab} \ne R_{ba}$, we break the symmetry. We can account for this asymmetry by explicitly including the rotation which takes one particle's vector into that of the other in the state description. Conventionally, however, this is not done. Hence, although such conventional state vectors in this common frame will still be symmetric under permutations (if they are uniquely defined) the distinguishing labels necessitated by the selection of one particle's independent frame relative to that of the other will introduce an exchange asymmetry.

This gives us the answer to the question we posed in section \ref{sec:asymcommon}: The asymmetry between $R_{ab}$ and $R_{ba}$ implies an asymmetry in any choice of common frame. However, if we always specify $R_a$ and $R_b$ with respect to the independent frames, such that $R_{ab}=R_{\mathbf{k}}(\pm\pi)$ then we have a means to handle this asymmetry in a common frame. By performing the appropriate rotations we can relate the exchange operation to permutation, compute the effect of exchanging the asymmetric distinguishing features and thereby obtain the exchange phase.

It is important to realize that the requirement $R_{ab}=R_{\mathbf{k}}(\pm\pi)$ for the relationship between the independent bisecting frames for a two-particle state vector applies even in the limit that $\hat{\mathbf{v}}_a$ and $\hat{\mathbf{v}}_b$ coincide. This is a crucial observation because, when $\hat{\mathbf{v}}_a=\hat{\mathbf{v}}_b$, it might be incautiously assumed that $R_{ab}$ was a null rotation. {\it It is because such an incautious assumption is made for the case of coincident spin quantization axes in the conventional construction, that the conventional exchange phase appears to be inexplicable as a result merely of permutation invariance}.

\section{Permutation Invariance In Momentum Space}\label{sec:momspace}
In section \ref{sec:USQ} we discussed uniqueness for single-particle state vectors of arbitrary momentum and spin. We now show how to use the prescription of the previous section to construct permutation-symmetric single-valued two-particle state vectors. We then show how the conventional construction is implicitly asymmetric and compute its exchange phase.

\subsection{Exchange Phase In Momentum Space}
Clearly, eqn. \ref{eqn:newwigdef} provides us with our desired uniqueness for single-particle state vectors. To define two-particle state vectors, we first of all take the symmetrized direct product vector for generalized spin quantization frames:
\begin{eqnarray}\label{eqn:gensymprod}
\lefteqn{|(Q_a,\mathbf{p}_a,s_a,m_a(R_a))_{B_a};(Q_b,\mathbf{p}_b,s_b,m_b(R_b))_{B_b}>}\nonumber\\
& = \alpha\ ( & 
|Q_a,\mathbf{p}_a,s_a,m_a(R_a)>_{B_a}|Q_b,\mathbf{p}_b,s_b,m_b(R_b)>_{B_b}\nonumber\\
& + & 
|Q_b,\mathbf{p}_b,s_b,m_b(R_b)>_{B_b}|Q_a,\mathbf{p}_a,s_a,m_a(R_a)>_{B_a}\ )\nonumber\\
& = & 
|(Q_b,\mathbf{p}_b,s_b,m_b(R_b))_{B_b};(Q_a,\mathbf{p}_a,s_a,m_a(R_a))_{B_a}>
\end{eqnarray}
where $R_a$ is the rotation which takes the base frame $B_a$ of particle $a$ into its spin quantization frame and likewise for $b$.

In section \ref{sec:USQ} we saw how to define canonical and helicity state vectors using particular choices of base frames. In the two particle case, where we may wish to relate the spin quantization frames to a common frame, we shall first look at how the selection of a particular choice of base frame for each particle can be made in a symmetric (order independent) way using one of the methods outlined in section \ref{sec:spatasym}. 

From sections \ref{sec:parallel} and \ref{sec:bisecting} and using the particle momenta $\hat{\mathbf{p}}_a$ and $\hat{\mathbf{p}}_b$ as the defining vectors $\hat{\mathbf{v}}_a$ and $\hat{\mathbf{v}}_b$, we see that there are two obvious choices for independent base frames. The first of these is the helicity frames (the parallel frames of section \ref{sec:parallel}):
\begin{eqnarray}\label{eqn:helframe}
\hat{\mathbf{z}}_c & = & \hat{\mathbf{p}}_c\nonumber\\
\hat{\mathbf{y}}_c & = & \hat{\mathbf{p}}_c\times\hat{\mathbf{p}}_o
\end{eqnarray}
(where $c=a$ and $o=b$ for the helicity frame of $a$ and $c=b$ and $o=a$ for the helicity frame of $b$). And the second choice is the momentum-bisecting frames:
\begin{eqnarray}\label{eqn:bisframe}
\hat{\mathbf{z}}_c & = & \hat{\mathbf{k}}\nonumber\\
\hat{\mathbf{y}}_c & = & \hat{\mathbf{p}}_c\times\hat{\mathbf{p}}_o
\end{eqnarray}
where $\mathbf{k}$ bisects $\hat{\mathbf{p}}_a$ and $\hat{\mathbf{p}}_b$.

In either case, and, in general, as we saw from section \ref{sec:genind}, the rotation $R_{ba}$, which relates the independent base frames, is then given by eqn. \ref{eqn:asymrot} to be $R_{\mathbf{k}}(\pm\pi)$.

This also means that if we are using a common spin quantization frame, then the rotations which take the independent base frames into the common frame must also be related by $R_{ba}$:
\begin{equation}\label{eqn:rotbtoa}
R_b = R_a.R_{ba} = R_a.R_{\mathbf{k}}(\pm\pi)
\end{equation} 
Hence, defining a single-valued and exchange-symmetric state vector in eqn. \ref{eqn:gensymprod} in a common canonical spin quantization frame depends  on the specification of a unique value for $R_{ba}$ that is unchanged when we permute the particles.

In general, of course, $R_a$ and $R_b$ can be independently defined to give arbitrary spin quantization frames for each particle.

\subsection{Exchange Symmetric Helicity States}
Taking the helicity frames as independent base frames, we then define the two-particle helicity state vector in a common frame of reference:
\begin{eqnarray}\label{eqn:xsymhel1}
\lefteqn{|Q_a,\mathbf{p}_a,s_a,\lambda_a;Q_b,\mathbf{p}_b,s_b,\lambda_b>^H}\nonumber\\
& = & |(Q_a,\mathbf{p}_a,s_a,\lambda_a(N))_H;(Q_b,\mathbf{p}_b,s_b,\lambda_b(N))_H>\nonumber\\
& = \alpha\ ( & |Q_a,\mathbf{p}_a,s_a,\lambda_a(N)>_H|Q_b,\mathbf{p}_b,s_b,\lambda_b(N)>_H\nonumber\\
& + & |Q_b,\mathbf{p}_b,s_b,\lambda_b(N)>_H|Q_a,\mathbf{p}_a,s_a,\lambda_a(N)>_H\ )\nonumber\\
& = \alpha\ ( &|Q_a,\mathbf{p}_a,s_a,\lambda_a>^H|Q_b,\mathbf{p}_b,s_b,\lambda_b>^H\nonumber\\
& + & |Q_b,\mathbf{p}_b,s_b,\lambda_b>^H|Q_a,\mathbf{p}_a,s_a,\lambda_a>^H\ )\nonumber\\
& = & |Q_b,\mathbf{p}_b,s_b,\lambda_b;Q_a,\mathbf{p}_a,s_a,\lambda_a>^H
\end{eqnarray}
and we see that because the base frame for each particle, is also its helicity frame, then the rotations which take the base frame into the spin quantization frames are null rotations for each particle.

If on the other hand, we had chosen the momentum-bisecting frames $M$ as our independent base frames, then we could still define the same permutation symmetric helicity state vector:
\begin{eqnarray}\label{eqn:xsymhel2}
\lefteqn{|Q_a,\mathbf{p}_a,s_a,\lambda_a;Q_b,\mathbf{p}_b,s_b,\lambda_b>^H}\nonumber\\
& = & |(Q_a,\mathbf{p}_a,s_a,\lambda_a(R))_M;(Q_b,\mathbf{p}_b,s_b,\lambda_b(R))_M>\nonumber\\
& = \alpha\ ( & |Q_a,\mathbf{p}_a,s_a,\lambda_a(R)>_M|Q_b,\mathbf{p}_b,s_b,\lambda_b(R)>_M\nonumber\\
& + & |Q_b,\mathbf{p}_b,s_b,\lambda_b(R)>_M|Q_a,\mathbf{p}_a,s_a,\lambda_a(R)>_M\ )
\end{eqnarray}
where $R$ takes the $M$ frame into the $H$ frame for each particle. Note that the variable $R$ can be dropped since it is implied by the values of the momenta and can be uniquely specified in an order independent way as a rotation by the angle $\theta$ in eqn. \ref{eqn:deftheta} (substituting $\mathbf{p}$ for $\mathbf{v}$) about the y-axis for each particle.

\subsection{Exchange Asymmetric Helicity States}
Now, suppose that, instead of choosing the base frames independently and symmetrically, we had chosen a common canonical frame to be the base frame for both particles. We still get the same permutation-symmetric helicity state vector:
\begin{eqnarray}\label{eqn:xsymhel3}
\lefteqn{|(Q_a,\mathbf{p}_a,s_a,\lambda_a(\bar{R}_a))_C;(Q_b,\mathbf{p}_b,s_b,\lambda_b(\bar{R}_b))_C>}\nonumber\\
& = \alpha\ ( & |Q_a,\mathbf{p}_a,s_a,\lambda_a(\bar{R}_a)>_C|Q_b,\mathbf{p}_b,s_b,\lambda_b(\bar{R}_b)>_C\nonumber\\
& + & |Q_b,\mathbf{p}_b,s_b,\lambda_b(\bar{R}_b)>_C|Q_a,\mathbf{p}_a,s_a,\lambda_a(\bar{R}_a)>_C\ )
\end{eqnarray}
where $\bar{R}_a = R(\hat{\mathbf{z}}\rightarrow \hat{\mathbf{p}}_a)$ and $\bar{R}_b = R(\hat{\mathbf{z}}\rightarrow \hat{\mathbf{p}}_b)$ must also obey 
\begin{equation}\label{eqn:rotbtoarev}
\bar{R}_b = R_{ab}.\bar{R}_a = R_{\mathbf{k}}(\pm\pi).\bar{R}_a
\end{equation} 

However, in this case, the cost of preserving permutation (and exchange) symmetry is the explicit inclusion of $\bar{R}_a$ and $\bar{R}_b$ in the state variables. Once again we see the advantage of using the independent symmetrically defined base frames such as $M$ and $H$, where this was not necessary.

We might think that one way to drop $\bar{R}_a$ and $\bar{R}_b$ is to assume a particular unambiguous value for them. For instance, if $\theta_a,\phi_a$ and $\theta_b,\phi_b$ are the angular co-ordinates of the momenta, defined in the normal range ($0\leq\theta_a,\theta_b<\pi$ and $0\leq\phi_a,\phi_b<2\pi$) then:
\begin{eqnarray}\label{eqn:defCtoH}
\bar{R}_a = R_{\mathbf{z}}(-\phi_a).R_{\mathbf{y}}(-\theta_a)\nonumber\\
\bar{R}_b = R_{\mathbf{z}}(-\phi_b).R_{\mathbf{y}}(-\theta_b)
\end{eqnarray}
However, it is easily seen that, because of eqn. \ref{eqn:rotbtoarev}, we cannot do this independently for both particles. Instead, we must adopt the strategy of defining this rotation for one particle and then defining the other relative to the first. This necessarily introduces an order dependence, which, for a single-valued state vector, we must make explicit (e.g. with the distinguishing superscripts ``1'' and ``2''):
\begin{eqnarray}\label{eqn:xasymhel}
\lefteqn{|(Q_a,\mathbf{p}_a,s_a,\lambda_a)^1;(Q_b,\mathbf{p}_b,s_b,\lambda_b)^2>^H}\nonumber\\
& = & |(Q_a,\mathbf{p}_a,s_a,\lambda_a(\bar{R}_a))_C;(Q_b,\mathbf{p}_b,s_b,\lambda_b(R_{12}.\bar{R}_a))_C>
\end{eqnarray}
where
\begin{equation}\label{eqn:defR12}
R_{12} = R_{\mathbf{k}}(\pm\pi).
\end{equation}
and it doesn't matter whether we choose the $+$ sign or the $-$ sign as long as we choose.

Now, under exchange, we find:
\begin{eqnarray}\label{eqn:xchghel}
\lefteqn{|(Q_b,\mathbf{p}_b,s_b,\lambda_b)^1;(Q_a,\mathbf{p}_a,s_a,\lambda_a)^2>^H}\nonumber\\
& = & |(Q_b,\mathbf{p}_b,s_b,\lambda_b(\bar{R}_b))_C;(Q_a,\mathbf{p}_a,s_a,\lambda_a(R_{12}.\bar{R}_b))_C>\nonumber\\
& = & |(Q_b,\mathbf{p}_b,s_b,\lambda_b(R_{ab}.\bar{R}_a))_C;(Q_a,\mathbf{p}_a,s_a,\lambda_a(R_{12}.R_{ab}.\bar{R}_a))_C>\nonumber\\
& = & |(Q_a,\mathbf{p}_a,s_a,\lambda_a(R_{12}.R_{ab}.\bar{R}_a))_C;(Q_b,\mathbf{p}_b,s_b,\lambda_b(R_{ab}.\bar{R}_a))_C>\nonumber\\
\end{eqnarray}
where the last step follows from the permutation symmetry (eqn. \ref{eqn:gensymprod}). We now have two cases. In the first case, $R_{ab} = R_{12}$ and we find
\begin{eqnarray}\label{eqn:xchghel1}
\lefteqn{|(Q_b,\mathbf{p}_b,s_b,\lambda_b)^1;(Q_a,\mathbf{p}_a,s_a,\lambda_a)^2>^H}\nonumber\\
& = & |(Q_a,\mathbf{p}_a,s_a,\lambda_a(R_{\mathbf{k}}(\pm 2\pi).\bar{R}_a))_C;(Q_b,\mathbf{p}_b,s_b,\lambda_b(R_{12}.\bar{R}_a))_C>\nonumber\\
& = & (-1)^{2s_a}\  |(Q_a,\mathbf{p}_a,s_a,\lambda_a)^1;(Q_b,\mathbf{p}_b,s_b,\lambda_b)^2>_H\nonumber\\
\end{eqnarray}
but in the second case, 
$R_{ab} = {R_{12}}^{-1} = R_{21}$, we find
\begin{eqnarray}\label{eqn:xchghel2}
\lefteqn{|(Q_b,\mathbf{p}_b,s_b,\lambda_b)^1;(Q_a,\mathbf{p}_a,s_a,\lambda_a)^2>^H}\nonumber\\
& = & |(Q_a,\mathbf{p}_a,s_a,\lambda_a(\bar{R}_a))_C;(Q_b,\mathbf{p}_b,s_b,\lambda_b(R_{\mathbf{k}}(\mp 2\pi).R_{12}.\bar{R}_a))_C>\nonumber\\
& = & (-1)^{2s_b}\  |(Q_a,\mathbf{p}_a,s_a,\lambda_a)^1;(Q_b,\mathbf{p}_b,s_b,\lambda_b)^2>_H\nonumber\\
\end{eqnarray}

It remains for us to clarify the difference between eqns. \ref{eqn:xchghel1} and \ref{eqn:xchghel2}. If both particles are fermions or both bosons, there is clearly no difference, since $(-1)^{2s_b} = (-1)^{2s_a}$. However, if one particle is a fermion and the other a boson, then we find a sign difference between the two versions of the exchange. However, this should be no surprise, since in this case the composite system is fermionic and the difference is due to an extra rotation of the {\it whole composite system} by $2\pi$ in one version of the exchange relative to the other, resulting in a sign change of $(-1)^{2s_a + 2s_b}$.

\subsection{The Two Particle Canonical State}\label{sec:2partcan}
Now let us turn our attention again to the canonical basis. We follow an analogous procedure to that used for the helicity basis. Using the independent helicity base frames $H$ defined by eqns. \ref{eqn:helframe}, we have:
\begin{eqnarray}\label{eqn:xsymbis}
\lefteqn{|(Q_a,\mathbf{p}_a,s_a,m_a(R_a))_H;(Q_b,\mathbf{p}_b,s_b,m_b(R_b))_H>}\nonumber\\
& = \alpha\ ( & |Q_a,\mathbf{p}_a,s_a,m_a(R_a)>_H|Q_b,\mathbf{p}_b,s_b,m_b(R_b)>_H\nonumber\\
& + & 
|Q_b,\mathbf{p}_b,s_b,m_b(R_b)>_H|Q_a,\mathbf{p}_a,s_a,m_a(R_a)>_H\ )\nonumber\\
& = &
|(Q_b,\mathbf{p}_b,s_b,m_b(R_b))_H;(Q_a,\mathbf{p}_a,s_a,m_a(R_a))_H> 
\end{eqnarray}
where $R_a$ and $R_b$ are the rotations which take the independent helicity frames for each particle into the common canonical frame. Clearly they are, again, related by eqn. \ref{eqn:rotbtoa}. Since $R_a$ and $R_b$ are unchanged by mere permutation, this state vector is permutation (and exchange) symmetric.

But if, as is conventionally done, we again wished to drop $R_a$ and $R_b$ from the notation, and assume unique implied values, then we face exactly the same problem as in the last section. Again we must adopt the strategy of defining the rotation for one particle and then defining the other relative to the first. And, again, this necessarily introduces an order dependence, which, for a single-valued state vector, we must again make explicit:
\begin{eqnarray}\label{eqn:xasymcan}
\lefteqn{|(Q_a,\mathbf{p}_a,s_a,m_a)^1;(Q_b,\mathbf{p}_b,s_b,m_b)^2>^C}\nonumber\\
& = & |(Q_a,\mathbf{p}_a,s_a,m_a(R_a))_H;(Q_b,\mathbf{p}_b,s_b,m_b(R_a.R_{21}))_H>
\end{eqnarray}
where
\begin{equation}\label{eqn:defR21}
R_{21} = R_{\mathbf{k}}(\pm\pi).
\end{equation}

Under exchange we have:
\begin{eqnarray}
\lefteqn{|(Q_b,\mathbf{p}_b,s_b,m_b)^1;(Q_a,\mathbf{p}_a,s_a,m_a)^2>^C}\nonumber\\
& = & |(Q_b,\mathbf{p}_b,s_b,m_b(R_b))_H;(Q_a,\mathbf{p}_a,s_a,m_a(R_b.R_{21}))_H>\nonumber\\
& = & |(Q_a,\mathbf{p}_a,s_a,m_a(R_a.R_{ba}.R_{21}))_H;(Q_b,\mathbf{p}_b,s_b,m_b(R_a.R_{ba}))_H>
\end{eqnarray}
Once again, choosing either $R_{ba} = R_{21}$ or $R_{ba} = R_{12}$, we find that the order dependence in the definition, creates an exchange asymmetry. Specifically, either
\begin{eqnarray}\label{eqn:xchgcan1}
\lefteqn{|(Q_b,\mathbf{p}_b,s_b,m_b)^1;(Q_a,\mathbf{p}_a,s_a,m_a)^2>^C}\nonumber\\
& = & (-1)^{2s_a}\  
|(Q_a,\mathbf{p}_a,s_a,m_a)^1;(Q_b,\mathbf{p}_b,s_b,m_b)^2>^C
\end{eqnarray}
or
\begin{eqnarray}\label{eqn:xchgcan2}
\lefteqn{|(Q_b,\mathbf{p}_b,s_b,m_b)^1;(Q_a,\mathbf{p}_a,s_a,m_a)^2>^C}\nonumber\\
& = & (-1)^{2s_b}\  
|(Q_a,\mathbf{p}_a,s_a,m_a)^1;(Q_b,\mathbf{p}_b,s_b,m_b)^2>^C
\end{eqnarray}

\section{The Exclusion Rules}
\label{sec:xrules}

Having defined both exchange symmetric and exchange asymmetric state vectors for both helicity and canonical frames of reference, it remains to determine the exclusion rules.

Exclusion rules arise whenever we can define a system that has some exchange asymmetry in it and we take the limit of identical individual quantum numbers.

There are two ways we can obtain state vectors that have exchange asymmetry. One example is when we define order dependent state vectors for pairs of separate single-particle states, as in eqns. \ref{eqn:xchghel1} and \ref{eqn:xchgcan1}. Obviously these are useful for fermions only, since the state vectors for boson pairs do not change sign under exchange. Another example occurs when we look at composite states with combined quantum numbers. Combining quantum numbers usually introduces an asymmetry because of the asymmetry in the coupling coefficients. This method is appropriate for both fermions and bosons and the simplest way to do this is to look at states of definite composite spin. The method can then be generalized to other quantum numbers.

\subsection{The Pauli Principle}
We start off by taking the limit of identical quantum numbers ($\mathbf{p}_a=\mathbf{p}_b = \mathbf{p}$, $Q_a = Q_b = Q$, $s_a = s_b = s$ and $m_a = m_b = m$) in the canonical basis. Then eqn. \ref{eqn:xchgcan1} becomes:
\begin{eqnarray}
\lefteqn{|(Q,\mathbf{p},s,m)^1;(Q,\mathbf{p},s,m)^2>^C}\nonumber\\
& = & (-1)^{2s}\  
|(Q,\mathbf{p},s,m)^1;(Q,\mathbf{p},s,m)^2>^C
\end{eqnarray}
which tells us that this state vector vanishes for fermions, giving us the Pauli Principle.

Similarly, in the helicity frame, the limit of identical quantum numbers means that the helicity frames coincide and we get exactly the same condition:
\begin{eqnarray}
\lefteqn{|(Q,\mathbf{p},s,\lambda)^1;(Q,\mathbf{p},s,\lambda)^2>^H}\nonumber\\
& = & (-1)^{2s}\  |(Q,\mathbf{p},s,\lambda)^1;(Q,\mathbf{p},s,\lambda)^2>_H\nonumber\\
\end{eqnarray}
The only difference being with the different spin quantization frame.

\subsection{The General Exclusion Rule For Composite Spin}
Now let us look at paired states of definite total spin $S$. Clearly, the quantization of the total spin requires a {\it single} unique choice of quantization frame.  To combine angular momentum, therefore, we need a common spin quantization frame for both particles, or, equivalently, must know how to relate the individual spin quantization frames to the total spin quantization frame. Obviously, the canonical basis is the most promising.

Similarly, unless we wish to contend with differing independent base frames, we must employ a system that has no need for explicit rotations which take the base frames into the canonical frame. The obvious choice is therefore the order dependent state vectors of eqn. \ref{eqn:xchgcan1}. Assuming two particles with the same spin $s_a = s_b = s$, The state vector for states of total spin S and third component M in the canonical frame are then given by:
\begin{eqnarray}
\lefteqn{|S,M:(Q_a,\mathbf{p}_a,s)^{1};(Q_b,\mathbf{p}_b,s)^{2}>^C}\nonumber\\
& = & \sum_{m_a,m_b} C^{ssS}_{m_am_bM}|(Q_a,\mathbf{p}_a,s,m_a)^{1};(Q_b,\mathbf{p}_b,s,m_b)^{2}>^C
\end{eqnarray}
where $C^{ssS}_{m_am_bM}$ are Clebsch-Gordon coefficients\cite{Rose} and the variables that precede the colon in the state vector on the left hand side are those which pertain to the composite state rather than the individual particles.

Since the $m_a$ and $m_b$ are summed over, we can interchange them in the terms being summed and, from the symmetry of the Clebsch-Gordon coefficients,
\begin{equation}
C^{ssS}_{m_am_bM} = (-1)^{S-2s} C^{ssS}_{m_bm_aM}
\end{equation}
we find
\begin{eqnarray}
\lefteqn{|S,M:(Q_a,\mathbf{p}_a,s_a)^{1};(Q_b,\mathbf{p}_b,s_b)^{2}>^C}\nonumber\\
& = & {1\over 2} \sum_{m_a,m_b} C^{ssS}_{m_am_bM}(|(Q_a,\mathbf{p}_a,s_a,m_a)^{1};(Q_b,\mathbf{p}_b,s_b,m_b)^{2}>^C\nonumber\\ 
& & + (-1)^{S -2s}|(Q_a,\mathbf{p}_a,s,m_b)^{1};(Q_b,\mathbf{p}_b,s,m_a)^{2}>^C )\nonumber\\
& = & {1\over 2} \sum_{m_a,m_b} C^{ssS}_{m_am_bM}(|(Q_a,\mathbf{p}_a,s_a,m_a)^{1};(Q_b,\mathbf{p}_b,s_b,m_b)^{2}>^C\nonumber\\
& & + (-1)^S |(Q_b,\mathbf{p}_b,s,m_a)^{1};(Q_a,\mathbf{p}_a,s,m_b)^{2}>^C ) 
\end{eqnarray}
the last step following from the exchange asymmetry of eqn. \ref{eqn:xchgcan1}. Now we see that, in the limit $\mathbf{p}_a=\mathbf{p}_b = \mathbf{p}$ and $Q_a = Q_b = Q$,
\begin{eqnarray}
\lefteqn{|S,M:(Q,\mathbf{p},s)^{1};(Q,\mathbf{p},s)^{2}>^C}\nonumber\\
& = & {1\over 2} (1 + (-1)^S)\sum_{m_a,m_b}C^{ssS}_{m_am_bM}
|(Q,\mathbf{p},s,m_a)^{1};(Q,\mathbf{p},s,m_b)^{2}>^C
\end{eqnarray}
from which we see that states of odd $S$ have vanishing state vectors. We therefore have the general exclusion rule, which applies to both bosons and fermions, that only states of even total spin $S$ are allowed when all other quantum numbers are identical. We would mention that this is simply a generalization of the Pauli Principle to particles of arbitrary spin, whether bosons or fermions, since, for spin $1\over 2$, the Pauli condition is exactly equivalent to the statement that the state $S=1$ is forbidden.

\subsection{General Exclusion Rules For Angular Momentum In The Center Of Mass Frame}

In their seminal paper on helicity states in the center of mass (CM) frame, Jacob \& Wick\cite{JW}(JW) show how to derive states of definite total angular momentum and definite helicity. They then show how the usual choice of exchange phase for identical particle states leads to a symmetry condition and give examples of exclusion rules for angular momentum states.

Our intention here is to revisit this issue and prove the general exclusion rule for states of definite $L,S$ in the CM frame.

First of all, we note that JW use an explicitly order dependent definition of the two-particle helicity state, when they rotate the spin quantization frame of particle ``2'' into that of particle ``1''. In their paper, they assume this to be a rotation by $\pi$ about the $y$-axis. However, to make the JW state vector single-valued, we relate it to our independent helicity state vector of eqn. \ref{eqn:xsymhel1}.  In this case, the rotation which takes the independent helicity frame of $b$ into that of $a$ is a rotation by $\pm\pi$ about $\mathbf{k}$ which, in the CM frame, is a rotation by $\pm\pi$ about the $z$-axis followed by a rotation by $\pm\pi$ about the $y$-axis.\footnote{Although a rotation $R_{\mathbf{y}}(\pi)$ alone, would also have resulted in the spin quantization of $b$ along the same axis as $a$, the final spin quantization frames in this case would have differed by a rotation by $\pm\pi$ about the $z$-axis. We think it is clear from JW that their intention was for the particles to share the same spin quantization frame. In actual fact, the reader can verify for themselves that either definition of the JW helicity two-particle state vector, with respect to our independent helicity state, will result in the same exclusion rule for $L$ and $S$.} Specifically, when $\mathbf{p}_a = \mathbf{p} = - \mathbf{p}_b$, we shall choose $R_{ba} = R_{\mathbf{y}}(\pi).R_{\mathbf{z}}(-\pi)$ to get:
\begin{eqnarray}\label{eqn:helJW}
\lefteqn{|\mathbf{p}_a:(Q_a,s_a,\lambda_a)^1;(Q_b,s_b,\lambda_b)^2>_{JW}}\nonumber\\
& = & |(Q_a,\mathbf{p}_a,s_a,\lambda_a(N))_H;(Q_b,\mathbf{p}_b,s_b,-\lambda_b(R_{\mathbf{y}}(\pi).R_{\mathbf{z}}(-\pi)
))_H>\nonumber\\
& = & (-1)^{s_b}
|Q_a,\mathbf{p}_a,s_a,\lambda_a;Q_b,\mathbf{p}_b,s_b,\lambda_b>^H
\end{eqnarray}

Under re-ordering, we have:
\begin{eqnarray}\label{eqn:symhelJW}
\lefteqn{|\mathbf{p}_b:(Q_b,s_b,\lambda_b)^1;(Q_a,s_a,\lambda_a)^2>_{JW}}\nonumber\\
& = & (-1)^{s_a}
|Q_b,\mathbf{p}_b,s_b,\lambda_b;Q_a,\mathbf{p}_a,s_a,\lambda_a>^H\nonumber\\
& = & (-1)^{s_a-s_b}
|\mathbf{p}_a:(Q_a,s_a,\lambda_a)^1;(Q_b,s_b,\lambda_b)^2>_{JW}
\end{eqnarray}

They then project out helicity partial wave states of definite total angular momentum J and third component M:
\begin{eqnarray}\label{eqn:pwJMab}
\lefteqn{|p,J,M:(Q_a,s_a,\lambda_a)^1;(Q_b,s_b,\lambda_b)^2>_{JW}}\\
& = & N_J\int{d\Omega 
{D^*}^J_{M\lambda_a-\lambda_b}(R(\mathbf{p}_a\rightarrow \mathbf{z})) |\mathbf{p}_a:(Q_a,s_a,\lambda_a)^1;(Q_b,s_b,\lambda_b)^2>_{JW}}\nonumber
\end{eqnarray}
where $p = |\mathbf{p}| = |\mathbf{p}_a| = |\mathbf{p}_b|$.

Similarly, interchanging $a$ and $b$, we have
\begin{eqnarray}\label{eqn:pwJMba}
\lefteqn{|p,J,M:(Q_b,s_b,\lambda_b)^1;(Q_a,s_a,\lambda_a)^2>_{JW}}\\
& = & N_J\int{d\Omega 
{D^*}^J_{M\lambda_b-\lambda_a}(R(\mathbf{p}_b\rightarrow \mathbf{z})) |\mathbf{p}_b:(Q_b,s_b,\lambda_b)^1;(Q_a,s_a,\lambda_a)2>_{JW}}\nonumber
\end{eqnarray}
and since
\begin{eqnarray}
R(\mathbf{p}_b\rightarrow \mathbf{z}) & = & R(\mathbf{p}_a\rightarrow \mathbf{z}).R_{ba}\\
{D^*}^J_{M\lambda_b-\lambda_a}(R(\mathbf{p}\rightarrow \mathbf{z}).R_{\mathbf{y}}(\pi).R_{\mathbf{z}}(-\pi)) & =  & (-1)^J{D^*}^J_{M\lambda_a-\lambda_b}(R(\mathbf{p}\rightarrow \mathbf{z}))\nonumber
\end{eqnarray}
then from eqn. \ref{eqn:symhelJW} we obtain the permutation (exchange) relation for the helicity partial wave state vectors:
\begin{eqnarray}\label{eqn:xasympwJM}
\lefteqn{|p,J,M:(Q_b,s_b,\lambda_b)^1;(Q_a,s_a,\lambda_a)^2>_{JW}}\nonumber\\
& = & (-1)^{J + s_a - s_b}
|p,J,M:(Q_a,s_a,\lambda_a)^1;(Q_b,s_b,\lambda_b)^2>_{JW}
\end{eqnarray}
and we see that, in the case of identical particles, then we obtain the same symmetry rule as eqn. ($47$) of JW, but without making any assumption (such as that which JW make) regarding the eigenvalues of the exchange operator since we have relied on permutation symmetry and single-valuedness instead.

To conclude this section, we note (again from JW) the connection with the $L-S$ coupling scheme:
\begin{eqnarray}
\lefteqn{|p,J,M,L,S:(Q_a,s_a)^1;(Q_b,s_b)^2>}\nonumber\\
& = & \left({{2J + 1}\over {2L+1}}\right)^{1\over 2}\sum_{\lambda_a \lambda_b} C^{LSJ}_{0\lambda \lambda} C^{s_as_bS}_{\lambda_a -\lambda_b \lambda}\nonumber\\
& & |p,J,M:(Q_a,s_a,\lambda_a)^1;(Q_b,s_b,\lambda_b)^2>_{JW}
\end{eqnarray}
where $\lambda = \lambda_a - \lambda_b$. In the limit of identical particles, and using the symmetry properties
\begin{eqnarray}
C^{ssS}_{\lambda_a -\lambda_b \lambda} & = & C^{ssS}_{\lambda_b -\lambda_a -\lambda}\nonumber\\
C^{LSJ}_{0\lambda \lambda} & = & (-1)^{J-L-S}C^{LSJ}_{0-\lambda -\lambda} 
\end{eqnarray}
we find, from eqn. \ref{eqn:xasympwJM}
\begin{eqnarray}
|p,J,M,L,S:(Q,s)^1,(Q,s)^2> = (-1)^{L+S} |p,J,M,L,S:(Q,s)^1,(Q,s)^2>
\end{eqnarray}
and hence obtain the generalized exclusion rule in the CM frame, that states of odd $L+S$ are excluded. Although not explicitly stated by Rose as a general rule, this rule is implicit in the many examples given in Rose's Chapter 12. The point of drawing the reader's attention to it is to clarify that the rule is the same for both fermions and bosons, once again showing that there is no essential physical difference between the two ``types'' of particles other than their spin and that both exhibit the same symmetry properties under identical particle permutation, leading to the same exclusion rule for composite states.

\section{Acknowledgment}

I would like to thank M. V. Berry, I. Duck, R. Mirman, J. Baez and D. H. Lyth for very useful, challenging and (sometimes) encouraging conversations which have been greatly beneficial to the presentation of this article. I would also like to acknowledge and thank my wife, Marianne Gontarz York for her joyfully uncritical encouragement and confidence, and my children, Alex and Keri, for their unconditional love.

\end{document}